\documentclass[a4paper,11pt]{article}
\pdfoutput=1 
\usepackage{jcappub} 
\usepackage{afterpage,booktabs,multirow} 
\usepackage{graphicx,subfig} 
\graphicspath{{./},{../plots/}}
\usepackage{amssymb,amsmath,bm} 
\usepackage[usenames,dvipsnames]{xcolor}
\usepackage[capitalise]{cleveref}
\crefname{equation}{Equation}{Equations}
\crefname{figure}{Figure}{Figures}
\crefname{table}{Table}{Tables}
\usepackage{expl3,xstring,xparse} 

\newcommand{\abs}[1]{\left| {#1} \right|}
\NewDocumentCommand\denv{o}{\delta_{\textrm{env}\IfValueT{#1}{,#1}}}
\newcommand{\fancyf}{\raisebox{-.2ex}{\text{{\large \textsl{f}}}}} 
\NewDocumentCommand\fr{o}{\IfValueTF{#1}{\abs{f_{\R 0}} = 10^{-#1}}{f(\R)}}
\NewDocumentCommand\given{}{\ensuremath{\, \middle| \,}}
\NewDocumentCommand\lcdm{}{\text{$\Lambda$CDM}}
\NewDocumentCommand\nueff{}{\nu_{\mathrm{eff}}}
\NewDocumentCommand\R{}{\mathcal{R}} 
\NewDocumentCommand\Renv{o}{R_{\textrm{env}}\IfValueT{#1}{ = #1/h\textrm{ Mpc}}}
\NewDocumentCommand\Senv{}{S_{\textrm{env}}}

\title{\boldmath The $f(\R)$ halo mass function in the cosmic web}
\author[a,1]{F. von Braun-Bates,\note{Corresponding author.}}
\author[a,b]{H. A. Winther,}
\author[a]{D. Alonso,}
\author[a,c]{and J. Devriendt}
\affiliation[a]{Astrophysics, University of Oxford, Denys Wilkinson Building, Keble Road, Oxford OX1 3RH, UK}
\affiliation[b]{Institute of Cosmology \& Gravitation, Dennis Sciama Building, University of Portsmouth, Portsmouth, PO1 3FX, UK}
\affiliation[c]{Observatoire de Lyon, UMR 5574, 9 avenue Charles Andre, F-69561 Saint Genis Laval, France}

\emailAdd{francesca.vonbraun-bates@physics.ox.ac.uk}
\emailAdd{hans.a.winther@physics.ox.ac.uk}
\emailAdd{david.alonso@physics.ox.ac.uk}
\emailAdd{julien.devriendt@physics.ox.ac.uk}

\abstract{An important indicator of modified gravity is the effect of the local environment on halo properties.  This paper examines the influence of the local tidal structure on the halo mass function, the halo orientation, spin and the concentration-mass relation. We use the excursion set formalism to produce a halo mass function conditional on large-scale structure. Our simple model agrees well with simulations on large scales at which the density field is linear or weakly non-linear. Beyond this, our principal result is that $f(\R)$ does affect halo abundances, the halo spin parameter and the concentration-mass relationship in an environment-independent way, whereas we find no appreciable deviation from \text{$\Lambda$CDM} for the mass function with fixed environment density, nor the alignment of the orientation and spin vectors of the halo to the eigenvectors of the local cosmic web. There is a general trend for greater deviation from \text{$\Lambda$CDM} in underdense environments and for high-mass haloes, as expected from chameleon screening.}

\begin{document}
\maketitle
\flushbottom

\section{Introduction} 
\label{sec:introduction}

The discovery of the accelerated expansion of the Universe \citep{Riess1998AJ....116.1009R,Perlmutter1999ApJ...517..565P} is one of the biggest puzzles in modern cosmology. Having accepted the Einstein equations as the correct description of the interaction between matter and geometry, one is left with no alternative but to modify either side of the equation in order to account for this acceleration.  On the one hand, modifying the stress-energy tensor necessitates the presence of an unknown substance dubbed dark energy; on the other, modifying the Einstein tensor require modifications to relativity at a more fundamental level.  This poses the question: should one of these choices be favoured over the other?  Dark energy---at its simplest a vacuum energy which acts in the form of an effective cosmological constant---is in excellent agreement with most observations so far, but suffers from some theoretical problems including fine-tuning \citep{Clifton:2011jh}.  Extended gravity theories---at their simplest adding extra complications to the Einstein-Hilbert action---have a secure theoretical motivation, but need to be carefully designed to satisfy observational constraints.

General Relativity has been tested to great precision on Earth and in the Solar System \citep{Berotti2003Natur.425..374B, Will2006LRR.....9....3W, 2015CQGra..32x3001B}.  Thus, if it is modified then some form of screening mechanism \citep{2004PhRvD..69d4026K, Khoury2010, 2015arXiv150404623K, Brax2012b} is required to hide the effects in these regimes. As we have tested general relativity only in high density regimes (relative to the cosmic mean), this naturally leads to an environmental dependence on the modifications of gravity.  In turn this would translate into an environmental dependence on observables \citep[e.g.][]{2010PhRvD..81j3002S, 2012ApJ...756..166W}. Thus the very screening which modified theories of gravity invoked to evade local, Solar System tests can be harnessed as a detection mechanism by examining different regions of the cosmic web. This can also be used to distinguish it from other dark energy scenarios.

This paper examines a variety of halo properties in the different geometric structures of the cosmic web, namely: the abundance of halos of different masses; the alignment of the halo spins and shapes with the local tidal structure; the halo spin parameter and the concentration-mass relation. We provide an excursion-set-based approximation to the {\it N}-body results for the halo mass function. This synthesises two, major existing results: the unconditional mass function in $f(\R)$ calculated by \cite{Lombriser:2013wta,Kopp:2013lea,2012MNRAS.421.1431L} and the classification of halo statistics in the cosmic web by e.g. \cite{1970Ap......6..320D,2009MNRAS.396.1815F,Alonso:2014zfa}. In the same vein as those papers, our aim is to build a semi-analytic model which combines an empirically-derived halo mass function in the unconditional case for \lcdm{} with a physical model derived using the tools of excursion set theory.  Such a model, if sufficiently accurate, can then be used to cheaply compute observables for quantities that otherwise would require expensive numerical simulations.

The structure of the paper is as follows: We summarise our chosen model of modified gravity in \cref{sub:review_of_modified_gravity}, before reviewing the excursion set theory in \cref{sub:the_excursion_set_theory_in_lcdm} (general relativity) and \cref{sub:the_excursion_set_theory_in_fR} (modified gravity) and by extending this to dependence on the cosmic web in \cref{sub:tidal_classification_of_the_cosmic_web}.  The {\it N}-body simulations against which we benchmark our semi-analytical model are described in \cref{sec:simulations_and_algorithms}. In \cref{sec:results} we compare the accuracy of our theoeretical model to {\it N}-body simulations (\cref{sub:volume_fractions,sub:multiplicity_functions_in_each_environment}) and we analyse deviations from \text{$\Lambda$CDM} for each of the halo properties (\cref{sub:multiplicity_functions_at_fixed_environment_density,sub:internal_halo_properties}).  We suggest possible generalisations and applications of this method in and conclude in \cref{sec:discussion}.

\section{Review of modified gravity} 
\label{sub:review_of_modified_gravity}

In this section we outline the differences between the modified gravity theory used in this paper and general relativity.  We motivate our choice of $f(\R)$ theory and describe some necessary conditions for which it is observationally viable.

An $f(\R)$ theory \citep{2010LRR....13....3D} adds a scalar function to the Einstein-Hilbert action for General Relativity.  It can be defined in the so-called Jordan frame via the action
\begin{equation} \label{eq:jordan_action}
S_J =\frac{1}{2}\int d^4 x\sqrt{-g}\, \left[\R+f(\R)\right] + \int d^4 x\sqrt{-g}\, {\mathcal{L}}_{\rm m}[\Phi_i,g_{\mu\nu}] \ ,
\end{equation}
where we have chosen units such that $8\pi G = 1$, the function $f(\R)$ is a general function of the Ricci scalar $\R$ and $\Phi_i$ denotes all matter fields.

For a modified gravity model of this kind to be observationally viable, it must exhibit the chameleon screening mechanism \cite{Noller:2013wca}. Via a field redefinition and a conformal transformation we can turn the Jordan frame action into an equivalent Einstein frame one \citep{2008PhRvD..78j4021B}
\begin{equation} \label{eq:einstein_action}
S_E =\frac{1}{2}\int d^4 x\sqrt{-\tilde{g}}\, \tilde{\R} + \int d^4 x\sqrt{-\tilde{g}}\,
\left[-\frac{1}{2}\tilde{g}^{\mu\nu}\tilde{\nabla}_{\mu}\phi \tilde{\nabla}_{\nu}\phi -V(\phi)\right]
+ S_{\text{matter}}[\Phi_i, e^{\beta \phi} \tilde{g}_{\mu\nu}]
\end{equation}
where a tilde denotes Einstein frame quantities and we have performed a conformal transformation
\begin{equation} \label{eq:conformal}
\tilde{g}_{\mu\nu} = e^{2 \omega} g_{\mu\nu} \quad \text{requiring} \quad e^{-2 \omega} \left( 1 + f_{\R} \right) = 1, \; \phi \equiv \frac{\omega}{\beta}
\end{equation}
where $\beta = \sqrt{1/6}$, $f_{\R} \equiv df(\R)/d{\R}$ and $V(\phi)$ is the scalar field self-interaction potential given by
\begin{equation} \label{eq:fr_potential}
    V(\phi) = \frac{1}{2}\frac{\R f_{\R}(\R) - f(\R)}{\left( 1 + f_{\R}(\R) \right)^{2}}
\end{equation}
The $f(\R)$ modification in the Jordan frame translates to a scalar-tensor theory of gravity in the Einstein frame where the scalar field $\phi$ is coupled to matter. In this formulation we can see that this model behaves as standard general relativity with the inclusion of a fifth-force mediated by the scalar field with coupling strength $1/3$, \textit{i.e.} in the unscreened limit, the gravitational force can be enhanced by a factor of up to $4/3$.  In the fully screened limit the gravitational force is not enhanced, \textit{i.e.} this modification does not manifest itself in dense environments or on small scales \cite{2014PhRvD..89b3523T}.  In order to exhibit this behaviour, the scalar field self-interation potential must satisfy a number of constraints (\textit{viz.} \cite{Noller:2013wca}), which again translates into constraints on the functional form of $f(\R)$ \citep{2007PhRvL..98m1302A}.

The explicit $f(\R)$ model we are working with in this paper is the Hu-Sawicki model of \cite{0705.1158}. This is a well studied model known to exhibit chameleon screening \cite{2010arXiv1011.5909K}. It is defined by
\begin{equation} \label{eq:fr_hu_sawicki}
f(\R) = - H_0^2\Omega_m \frac{\left(c_1 \left(\frac{\R}{m^2}\right)^{n_{f(\R)}}\right)}{1 + c_2 \left( \frac{\R}{m^2}\right)^{n_{f(\R)}}}
\quad \text{where $c_1 = \frac{6\Omega_{\Lambda}}{\Omega_m}c_2 $}
\end{equation}
Rather than using $c_2$ as the free parameter we will instead express it in terms of $f_{\R 0}$, the value of $f_{\R}$ in the cosmological background evaluated at $z=0$, using
\begin{equation}
|f_{\R 0}| = n_{f(\R)} \frac{|c_1|}{c_2^2}\left(\frac{\Omega_m}{3(\Omega_m + 4\Omega_\Lambda)}\right)^{1+n_{f(\R)}}
\end{equation}
In this paper we will only consider $n_{f(\R)} = 1$ and $|f_{\R 0}| = 10^{-5}$.

To see how screening works, let us consider a top-hat overdensity of radius $R_{\textrm{TH}}$ and mass $M_{\rm TH}$.  The Newtonian potential of the overdensity is $\Phi_N = \frac{GM_{\rm TH}}{R_{\rm TH}}$. As shown in \cite{2008PhRvD..78j4021B} the gravitational force on a test-mass of mass $m$ outside the top-hat is approximately given by
\begin{align}
F = \frac{GM_{\rm TH}m}{r^2}\left(1 + \frac{1}{3}\frac{\Delta R}{R_{\rm TH}}\right)
\end{align}
where the screening factor $\frac{\Delta R}{R_{\rm TH}}$ is given by
\begin{equation}\label{eq:screening}
\frac{\Delta R}{R_{\textrm{TH}}} = \min\left\{\frac{3|f_{\R}^{\rm TH} - f_{\R}^{\rm env}|}{2\Phi_N} ,1\right\}
\end{equation}
where $f_{\R}^{\rm TH} = f_{\R}(\rho_{\rm TH})$ and $f_{\R}^{\rm env} = f_{\R}(\rho_{\rm env})$ are the scalar field values inside and outside the body respectively. For the Hu-Sawicki model the scalar field value in a region density $\rho_m$ and time (scale-factor) $a$ can be estimated to be
\begin{align}
f_{\R}(\rho_m) = f_{\R 0} \left(\frac{1  + \frac{4\Omega_{\Lambda 0}}{\Omega_{m0}} }{\frac{\rho_m}{\rho_{m0}} + \frac{4\Omega_{\Lambda 0}}{\Omega_{m0}} }\right)^{n_{f({\R})}+1}
\end{align}
where $\rho_{m0} = 3\Omega_{m0}H_0^2M_{\rm Pl}^2$ is the average matter density in our Universe at the present time. When the overdensity is massive ($\frac{1}{\Phi_N}$ is very small) or is located in a very dense environment ($|f_{\R}^{\rm TH} - f_{\R}^{\rm env}|$ is very small) then $\frac{\Delta R}{R_{\textrm{TH}}} \ll 1$ and the fifth-force is screened. In contrast, when the overdensity is not massive (so $\frac{1}{\Phi_N}$ is very large) then $\frac{\Delta R}{R_{\textrm{TH}}} \approx 1$ and the force is $4/3$ the value of the Newtonian prediction.  Thus we see that the modification to gravity is sensitive to both the halo mass and the environment density.

A more accurate expression for the screening factor, which we will use in this paper, was derived in \cite{Lombriser:2013wta} from results in \cite{KhouryWeltman,2012MNRAS.421.1431L}. The gravitational force on a test-particle at the surface of a top-hat overdensity of radius $R_{\textrm{TH}}$ collapsing in an expanding background is given by
\begin{equation} \label{eq:feff_newtonian}
F = \frac{GM_{\rm TH}m}{r^2}\left[1 + \frac{1}{3}F_{\rm eff}\right]
\end{equation}
where
\begin{subequations} \label{eq:feff_fr}
\begin{align}
    F_{\textrm{eff}}(a, R_{\rm TH},\rho_{\rm TH},\rho_{\rm env}) &= \frac{1}{3} \left[ 3 \left( \frac{\Delta R}{R_{\textrm{TH}}} \right)
    - 3 \left( \frac{\Delta R}{R_{\textrm{TH}}} \right)^{\!2}
    + \left( \frac{\Delta R}{R_{\textrm{TH}}} \right)^{\!3} \right] \\
    \frac{\Delta R}{R_{\textrm{TH}}} &=\min\left\{ \frac{\rho_{m0}}{\Omega_{m0}(R_{\rm TH}H_0)^2 \rho_{\rm TH}} \left| f_{\R}(\rho_{\rm TH}) - f_{\R}(\rho_{\rm env})\right|,1\right\}
\end{align}
\end{subequations}
The background cosmology of this model is always very close to $\Lambda$CDM as long as $|f_{\R 0}| \ll 1$ (which is required to satisfy local gravity constraints). We have three regimes: on cosmological scales, the background solution mimics \text{$\Lambda$CDM}; on local scales in high-density regions, the modification is screened in order to evade tight, Solar System constraints; on local scales in low-density regions the modifications of gravity are in full effect.


\section{Halo abundances in cosmology} 
\label{sec:halo_abundances_in_cosmology}

In this section we outline the excursion set in general relativity (\cref{sub:the_excursion_set_theory_in_lcdm}) and then discuss the modifications induced by modified gravity (\cref{sub:the_excursion_set_theory_in_fR}).  Finally we present the key integral of this paper, which incorporates the dependence on the cosmic web (\cref{sub:tidal_classification_of_the_cosmic_web}). In particular, we choose the Peacock mass function \cref{eq:JAP_uncond_cmf} as our unconditional model in \lcdm{} and we discuss which aspects of the excursion set theory in modified gravity are compatible with our model for the cosmic web. In this way, we ensure that our base model is an excellent fit to simulations, while also retaining the semi-analytical nature of the model.

\subsection{The excursion set theory in \text{$\Lambda$CDM}} 
\label{sub:the_excursion_set_theory_in_lcdm}

The \textit{Ansatz} of the excursion set approach is to relate the fraction of the density field $\delta$ above a certain critical density $\delta_{c}$ to the cumulative fraction $F(>M)$ of mass contained in haloes above mass $M$.

The overdensity field $\delta$ is obtained from the Gaussian fluctuations in the fractional, linearly-evolved overdensity smoothed over a scale $R$ defined by a given Fourier-space window function $W(kR)$. In what follows we will relate the smoothing scale to a mass scale $M$ defined as the mass encompassed by the kernel $W(kR)$ in a homogeneous Universe.

The fluctuations of the overdensity field have some variance $\sigma^2$, and we define the variable $\nu\equiv\delta/\sigma$. In what follows we will refer to the variance at redshift zero $S\equiv\sigma^2(M,z)/D(z)^2$ (where $D(z)$ is the linear growth factor normalised to $D(z=0)=1$) as the {\sl resolution}, which is related to the linear matter power spectrum as:
\begin{equation} \label{eq:variance}
    S(M)\equiv\frac{1}{2 \pi^2} \int_{0}^{\infty} dk\, k^2\,P(k,z=0)\,|W(kR)|^2.
\end{equation}

The foundation of the excursion set formalism is the relation of the linear overdensity field in real space collapsing to a halo of mass $M$ to
trajectories in the density field being absorbed by a critical density at resolution $S$. As shown in \cite{1991ApJ...379..440B,1993MNRAS.262..627L}, where they assume top-hat window functions in $k$-space, this problem reduces to finding the fraction of random walks in the plane $(\delta,S)$ that are absorbed by the collapse overdensity $\delta_c$ on masses larger than $M$. For Gaussian random fields this has a simple solution given by the Press-Schechter mass function \cite{1974ApJ...187..425P}:
\begin{equation} \label{eq:ps_erfc_integral}
    F(>M) = 2 \int_{\delta_{c}}^{\infty} P(\delta(S))\,d\delta=\textrm{erfc} \left(\frac{\nu_h}{\sqrt{2}} \right),
\end{equation}
where $\nu_h\equiv\delta_c/\sqrt{S}$.

The qualitatively-useful Press-Schechter distribution of \cite{1974ApJ...187..425P} has long been supplanted---at least in general relativity---by a variety of empirical fits to {\it N}-body simulations (e.g. see \cite{Sheth:2001dp,2006ApJ...646..881W,2007MNRAS.374....2R,2008ApJ...688..709T}), often preserving the assumption of \textsl{universality} of the collapsed mass fraction. In this context, expressing $\mathrm{d} F(>M) / \mathrm{d} \ln \nu$ in units of $\nu_h$ makes the resulting expression independent of the choice of cosmological parameters \cite{Zentner:2006vw}. In this paper, we adopt the Peacock mass function\footnote{This simple functional form matches the Warren et al. fit \cite{2006ApJ...646..881W} to the mass function to sub-percent accuracy.} \cite{Peacock2007}
\begin{equation} \label{eq:JAP_uncond_cmf}
    F(>M) =
    \frac{\exp\left( -c \nu_h^2\right)}{1 + a\,\nu_h^b}
    \quad \text{where} \quad \begin{cases}
    a &= 1.529 \\
    b &= 0.704 \\
    c &= 0.412
    \end{cases}
\end{equation}
Instead of the collapsed mass fraction (Eq. \ref{eq:JAP_uncond_cmf}), we will use its differential, known as the multiplicity function\footnote{Note that some authors define the multiplicity function as the derivative with respect to $\sigma$ rather than $\ln M$.  Moreover, although $f(M)$ and $\fancyf{}(S)$ are related, they do not denote the same functions, despite the same letter being conventionally used for both.}:
\begin{equation} \label{eq:multiplicity}
    f(M) \equiv \frac{d F(<M)}{d{\ln M}}
    = \fancyf (S) \abs{\frac{d{S}}{d{\ln M}}}
    = - \langle \fancyf \left( \delta_{c} (S), S \given \denv, \Senv \right) \rangle_{\mathrm{env}} \frac{d{S}}{d{\ln M}}
\end{equation}
where we have defined the first-crossing distribution $\fancyf(S)$, the probability that the random walk will be absorbed by the barrier at resolution $S$. We will see in the next section that this becomes much more important in modified gravity than general relativity.
\begin{figure}
  \centering
  \includegraphics[width=0.7\textwidth]{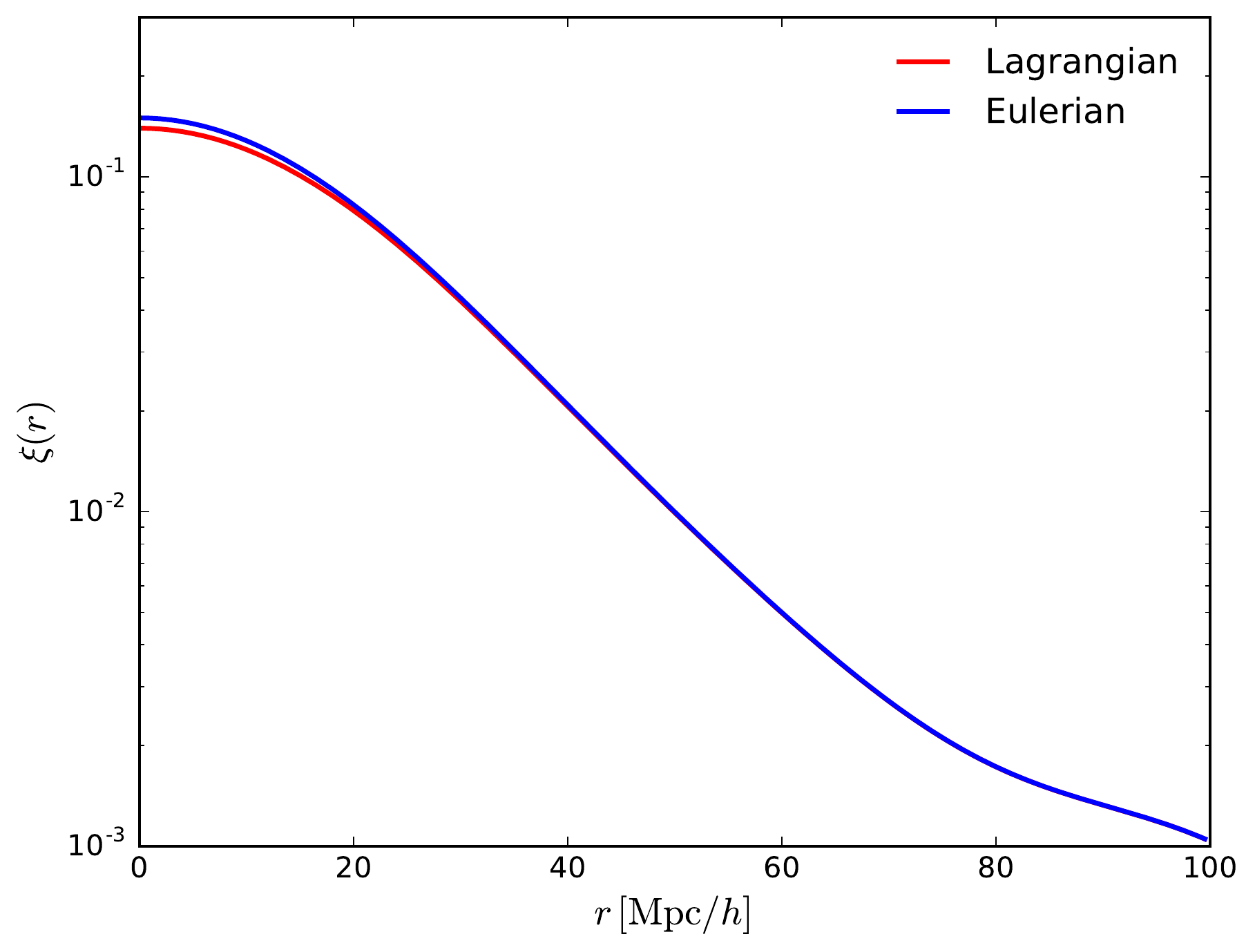}
  \caption{Two-point correlation function of the Lagrangian (red) and Eulerian (blue) matter density fields, the latter approximated by a lognormal transformation of the former. The results are shown for a Gaussian smoothing scale $R_{\rm env}=10\,{\rm Mpc}/h$. The relative difference between both curves is smaller than 7\% on all scales.}  \label{fig:evsl}
\end{figure}

We finish this section by noting that, in a strict sense, when relating the excursion-set predictions to simulated or real data, the scale of the environment $S_{\rm env}$ should correspond to its Lagrangian scale (i.e. in the initial conditions) instead of its Eulerian size, and as pointed out by \cite{2012MNRAS.426.3260L,2012MNRAS.425..730L} these effects could be relevant for modified-gravity theories. However, for the environment scales studied in this work ($R_{\rm env}\geq10\,{\rm Mpc}/h$), we expect these two quantities to be very similar, the evolution of the density field since on such large scales is close to self-similar. We can verify this with a quick test: the effect of the Eulerian evolution can be roughly captured by making a log-normal transformation on the linear (Lagrangian) density field \cite{Coles01011991}. Then, since the shape of density profile around any point in either the Lagrangian or Eulerian fields is, on average, determined by the correlation function of the field, we can explore the effects of this distinction by studying the correlation function in either case. Under a lognormal transformation, the correlation function of the log-normalized (i.e. Eulerian) field would be given by $\xi_{\rm Eu}(r)=\exp[\xi_{\rm Lag}(r)]-1$, where $\xi_{\rm Lag}$ is the correlation function of the linearized (i.e. Lagrangian) field. Figure \ref{fig:evsl} shows both correlation functions for the smallest smoothing scale explored in this work ($R=10\,{\rm Mpc}/h$), and shows the negligible effect of ignoring the true Eulerian scale of the environment. This completes the excursion set theory in general relativity.

\subsection{The excursion set theory in $f(\R)$} 
\label{sub:the_excursion_set_theory_in_fR}

\begin{figure}[htbp]
    \begin{center}
        \subfloat[]{\label{subfig:plot_deltac_mass}\includegraphics[keepaspectratio,width=0.7\textwidth]{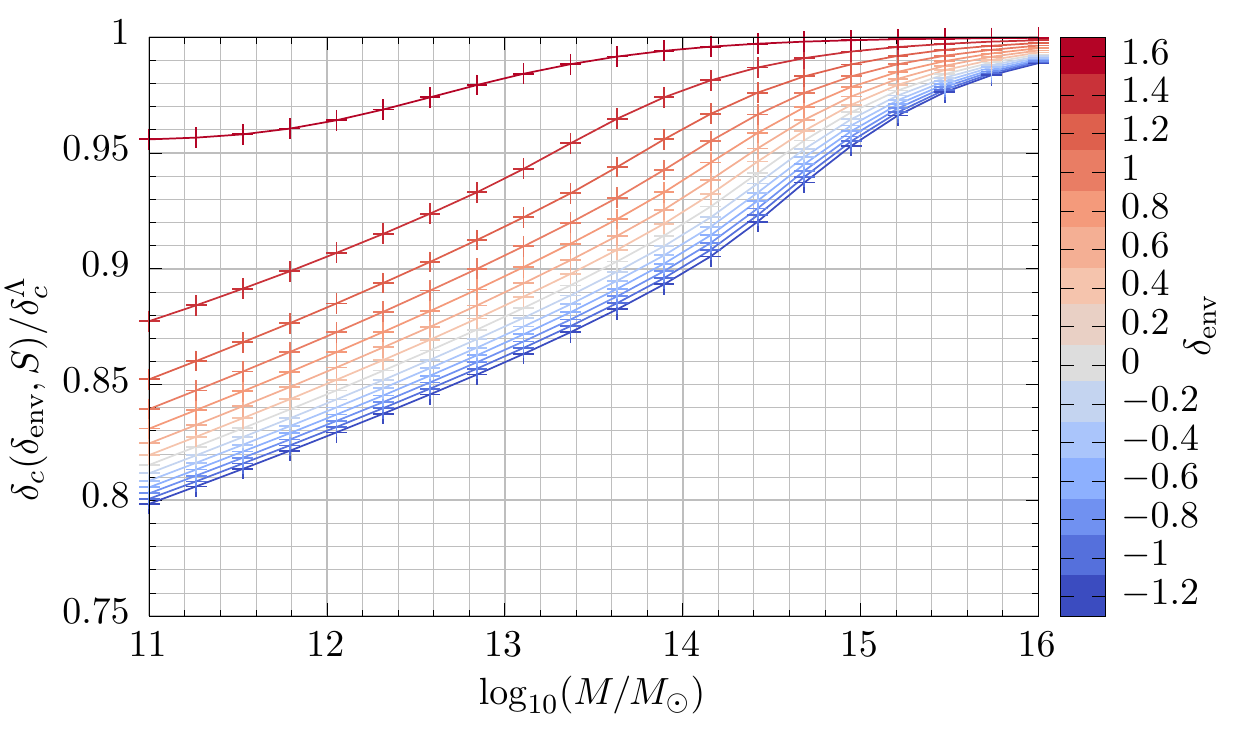}} \\
        \subfloat[]{\label{subfig:plot_deltac_denv}\includegraphics[keepaspectratio,width=0.7\textwidth]{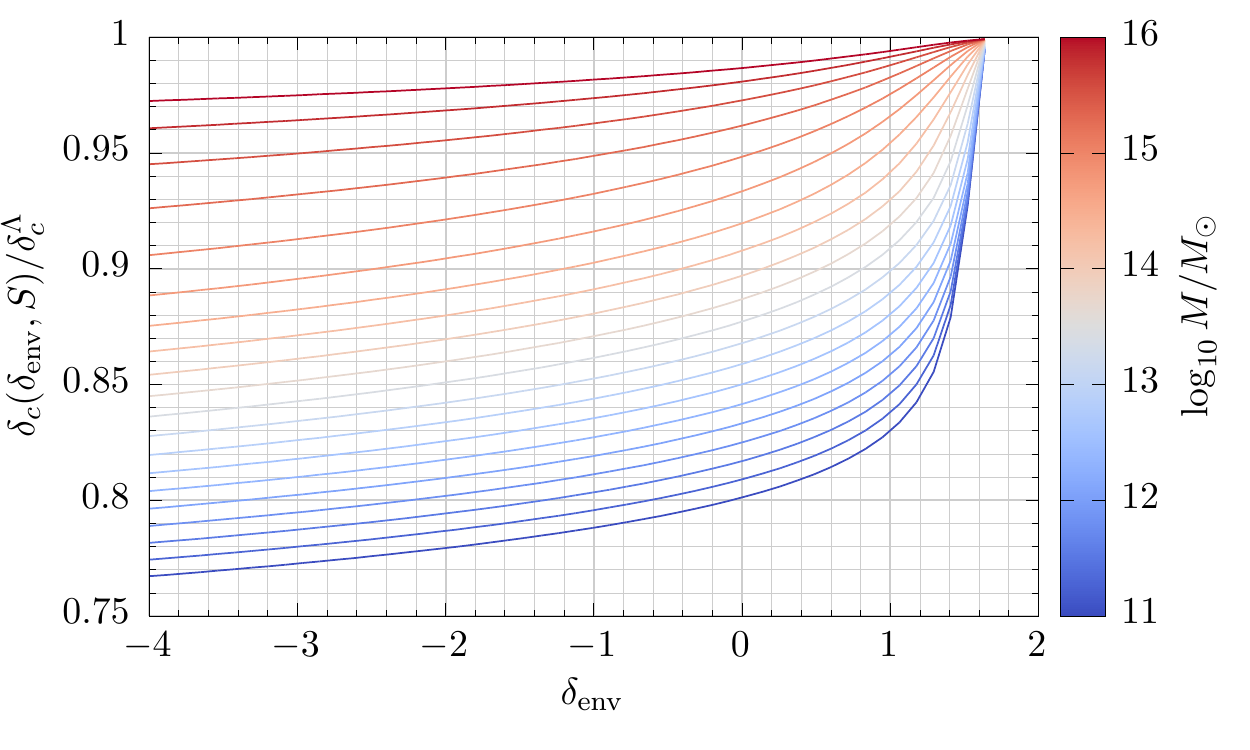}}
    \end{center}
    \caption{The $f(\R)$ barrier $\delta_{c}(z,\denv,M)$ compared to the \text{$\Lambda$CDM} barrier $\delta^{\Lambda}_{c}(z)$.}
\end{figure}

\begin{figure}
	\begin{center}
		\includegraphics[keepaspectratio,width=0.7\textwidth]{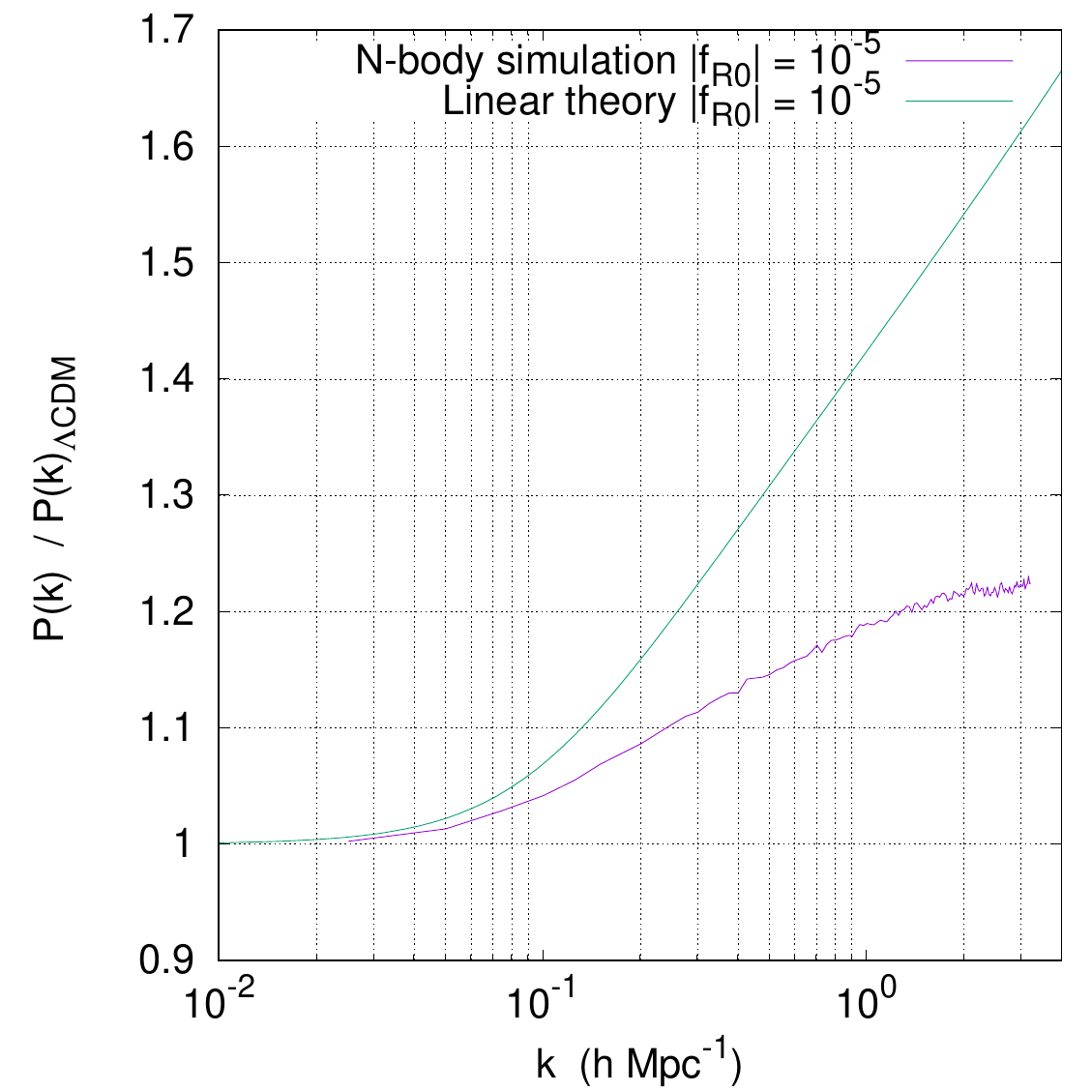}
	\end{center}
	\caption{The ratio of the $f(R)$ power spectrum to that of \lcdm{} for the linear (purple) and non-linear (green) case.  The effect of screening is clearly visible at increasingly large $k$: the non-linear result is suppressed, while the linear result increases rapidly.}
	\label{fig:pofk_f5}
\end{figure}

We now summarise the differences between excursion set theory in $f(\R)$ gravity compared to \text{$\Lambda$CDM}.  This section presents a formula for the overdensity required for collapse $\delta_{c}(M,z,f_{\R 0},\delta_{\rm env})$. The main difference in modified gravity is that $\delta_c$ depends both on mass, environment density and redshift apposed to just depending on redshift as in $\Lambda$CDM. We then discuss the consequences for calculating the first-crossing distribution $f(S)$ and state which of these we will include in our model.

The collapse of the environment surrounding the halo is equivalent to collapse in general relativity.  Let us assume that the initial overdensity is a spherical top-hat in Eulerian space.  We can utilise the resulting axisymmetry to simplify the gravitational collapse equation to:
\begin{equation} \label{eq:d_nonlinear_gr}
    0 = \frac{d^2{y}}{d \ln a^{2}}
    + \left(2 - \frac{3}{2} \Omega_{m}(a) \right) \frac{d{y}}{d \ln a}
    + \frac{1}{2}  \Omega_{m}(a) \left( \frac{1}{y^{3}} - 1 \right) y
\end{equation}
where $y(a)$ is the ratio of the physical radius of the halo $R_{\rm TH}(a)$ to the physical radius of the filter $a(t)R$ \cite{2012MNRAS.421.1431L}.  The initial conditions are obtained by mass conservation, which forces the equality \cite{Lombriser:2013wta}:
\begin{equation} \label{eq:mass_conservation}
   M = \frac{4}{3} \pi \rho_{m0} a_{i}^{3} R^{3} = \frac{4}{3} \pi \rho r_{i}^{3}
   \implies y_{i} = 1 - \frac{\delta_i}{3} \quad \text{and} \quad \frac{d{y}_{i}}{d{\ln a}} = - \frac{\delta_i}{3}.
\end{equation}
An important corollary of \cref{eq:mass_conservation} is that $\delta = y^{-3} - 1$, which contributes the nonlinearity in the final term of \cref{eq:d_nonlinear_gr}.  Now we have an expression for the halo density in general relativity and the environment density in modified gravity.

It is possible to show that, in $f(\R)$, Eq. \ref{eq:d_nonlinear_gr} takes the form:
\begin{equation} \label{eq:d_nonlinear_gen}
    0 = \frac{d^2{y}}{d \ln a^{2}}
    + \left(2 - \frac{3}{2} \Omega_{m}(a) \right) \frac{d{y}}{d \ln a}
    + \left( 1 +  F_{\mathrm{eff}} \right) \frac{1}{2}  \Omega_{m}(a) \left( \frac{1}{y^{3}} - 1 \right) y,
\end{equation}
where $F_{\mathrm{eff}} = F_{\mathrm{eff}}(a, R_{\rm TH}, \rho_{\rm TH}, \rho_{\rm env})$ was introduced in \cref{eq:feff_fr}, with $R_{\rm TH} = y a R$, $\rho_{\rm TH} = \rho_{m0} a^{-3} y^{-3}$. Since $F_{\mathrm{eff}}  \ll 1$ at early times, we can use the same initial conditions as we used for $\Lambda$CDM.  This is the general form for modified gravity collapse of a spherical top-hat.  Substituting $F_{\mathrm{eff}}$ from \cref{eq:feff_fr} determines the \textsc{ode} for collapse of the halo in modified gravity.

How does this affect the excursion set formalism? After solving Eqs. \cref{eq:d_nonlinear_gr,eq:d_nonlinear_gen} we can derive values for $\delta_{c}$ in general relativity and modified gravity, and for the starting point of the random walk $\denv$.  So far we have neglected the additional complications that the resolution $S$ is sensitive to $f(\R)$ via the power spectrum and that the collapse density is now a function of $S$.  We address these now.

We have chosen to use the linear $P(k)$ computed in the $\Lambda$CDM model in \cref{eq:variance} when computing the $f(\R)$ gravity predictions. The same approach was taken in \cite{2016JCAP...12..024C}. The reason for this choice is that linear theory massively overestimates the clustering in modified gravity theories such as $f(\R)$: \cref{fig:pofk_f5} shows that the ratio of the linear power-spectrum in $\fr[5]$ to the linear power-spectrum in $\Lambda$CDM shows a much larger deviation than the ratio of the corresponding non-linear power spectra obtained from simulations. The reason is that screening is not present in linear theory.  One can easily construct modified gravity models of this type where non-linear results for the power spectrum and the mass function can be arbitrarely close to $\Lambda$CDM while at the same time having a linear power-spectrum that deviates significantly from $\Lambda$CDM. In such scenarios excursion set theory is not able to give results that agree with simulations if the modified gravity linear power spectrum is used in the calculations. We have checked that this gives result that are in better agreement with simulations by comparing the results using the full linear power-spectrum. An alternative would be to consider a compromise approach where we use the linear $\Lambda$CDM power-spectrum corrected with a boost-factor $P_{f(\R)}(k) /P_{f(\Lambda\rm CDM}(k) $ computed from simulations, however this would require explicit simulations which goes against the appeal of using the excursion set approach---which is to extract observables without having to perform expensive numerical calculations.

The barrier density is no longer flat: $\delta_{c}(S, \denv)$ is a montotonically-decreasing (-increasing) function of $S$ ($M$; \cref{subfig:plot_deltac_mass}), and the peak-backgound split $\delta_{c} - \denv$ is a montotonically-decreasing function of $\denv$(\cref{subfig:plot_deltac_denv}).  Compared to $\Lambda$CDM, we expect haloes to form at higher masses (earlier in the random walk) and to have more in low-density regions (where the fifth-force potential is unscreened).  For a moving barrier like this, and under the assumption that trajectories in $(\delta,S)$ take uncorrelated steps, the diffuson equation admits a solution \cite{2011PhRvD..83f3511P}:
\begin{subequations} \label{eq:volterra_all}
\begin{align}
    g \left( S \given S_{\mathrm{env}} , \delta_{\mathrm{env}} \right)
    &= h(S) + \int_{S_{\mathrm{env}}}^{S} \! d{x} \; k(S,x) \, g \left( x \given S_{\mathrm{env}} , \delta_{\mathrm{env}} \right) \label{eq:volterra_fcd} \\
    k(S,x) &= \left[ \frac{\delta_{c}(S) - \delta_{c}(x)}{S - x} - 2\frac{d{\delta_{c}(S)}}{d{S}} \right]
    \frac{1}{\sqrt{2\pi (S-x)}} \exp\left\{ -\frac{(\delta_{c}(S) - \delta_{c}(x))^{2}}{2 (S-x)} \right\} \\
    h(S) & = \left[ \frac{\delta_{c}(S) - \delta_\mathrm{env} }{S - S_{\mathrm{env}} } - 2\frac{d{\delta_{c}(S)}}{d{S}} \right]
    \frac{1}{\sqrt{2\pi (S - S_{\mathrm{env}})}} \exp \left\{ -\frac{(\delta_{c}(S) - \delta_{\mathrm{env}})^{2}}{2(S - S_{\mathrm{env}})} \right\}
\end{align}
\end{subequations}
It is straightforward to show that this integral equation has an analytical solution for a constant barrier \cite{Zhang:2005ar,2011PhRvD..83f3511P}, including the constant case $\delta_{c}^{\Lambda} \approx 1.676$ which is the \text{$\Lambda$CDM} solution.
Once the first crossing distribution is found after solving this equation, the procedure would be to marginalise over the possible values of the environmental density. This procedure is numerically complicated and, since its derivation is based on purely uncorrelated random walks, which do not apply to the environment definitions used in this work, we took a simpler approach, based on including the effects of $f(\R)$ simply by substituting the expression for the mass- and environment-dependent collapse threshold $\delta_c(S,\denv)$ in the $\Lambda$CDM result described in the next section.

\subsection{Tidal classification of the cosmic web} 
\label{sub:tidal_classification_of_the_cosmic_web}

Now we build on the unconditional Press-Schechter result in \cref{sub:the_excursion_set_theory_in_lcdm} to find the equivalent result dependent upon the local environment. As done in e.g. \cite{2009MNRAS.396.1815F,Alonso:2014zfa}, we characterise the environment in terms of the properties of the local tidal tensor $\mathsf{T}_{ij}$, defined as the the Hessian of the normalised, Newtonian potential $\phi$ smoothed with a given kernel of size $R_{\rm env}$
\begin{equation} \label{eq:tidal_tensor}
    \mathsf{T}_{ij} = \frac{\partial_{i} \partial_{j} \phi}{4 \pi G \bar\rho}.
\end{equation}
The environment is then classified into one of four different classes, depending on the values of the three eigenvalues  of ${\sf T}$, $\lambda_{1} \geq \lambda_{2} \geq \lambda_{3}$, in relation to a given eigenvalue threshold. We thus define voids, sheets, filaments and knots as regions in which $\alpha \in \left\{ 0,1,2,3 \right\}$ eigenvalues are above the threshold respectively. Since the tidal tensor quantifies the direction and intensity of the local tidal forces, this classification thus informs about the number of dimensions in which extended objects are contracted or stretched. We will call these environment types ``elements'' of the cosmic web.

For convenience, let $S_{\rm env}$ be the variance of the overdensity field convolved with the kernel defining the environment, and let us introduce the three useful quantities:
\begin{subequations} \label{eq:eigenvalues_all}
\begin{align}
    \nu_{e} &= \left( \lambda_{1} + \lambda_{2} + \lambda_{3} \right) / \sqrt{\Senv}, \\
    \theta  &= \frac{1}{2} \left( \lambda_{1} - 2 \lambda_{2} + \lambda_{3} \right) / \sqrt{\Senv}, \\
    \rho  &= \frac{1}{2} \left( \lambda_{1} - \lambda_{3} \right) / \sqrt{\Senv}.
\end{align}
\end{subequations}
The classification scheme outlined above can be cast in terms of the values of these quantities as:
\begin{subequations} \label{eq:nueff_integral_limits}
\begin{align}
    \nu_{e} - \nu_{\mathrm{th}} &\in \left[ f_{1}\left( \rho, \theta \given \alpha \right) \, , \, f_{2}\left( \rho, \theta \given \alpha \right) \right]
    \quad \text{where $\nu_{\mathrm{th}} = 3 \frac{\lambda_{\text{th}}}{\sqrt{\Senv}}$} \\
    f_{1}\left( \rho, \theta \given \alpha \right) &=
    \begin{cases}
        -\infty              &\alpha = 0 \quad \text{(voids)} \\
        -3 \rho - \theta  &\alpha = 1 \quad \text{(sheets)} \\
        2 \theta            &\alpha = 2 \quad \text{(filaments)} \\
        3 \rho - \theta   &\alpha = 3 \quad \text{(knots)}
    \end{cases} \\
    f_{2}\left( \rho, \theta \given \alpha \right) &=
    \begin{cases}
        -3 \rho - \theta    &\alpha = 0 \quad \text{(voids)} \\
        2 \theta              &\alpha = 1 \quad \text{(sheets)} \\
        3 \rho - \theta     &\alpha = 2 \quad \text{(filaments)} \\
        \infty                 &\alpha = 3 \quad \text{(knots)}
    \end{cases}
\end{align}
\end{subequations}
where $\lambda_{\text{th}}$ is the eigenvalue threshold, and $\alpha$ is the number of eigenvalues above it.

It can be proven that the probability distribution for the environmental tidal field in terms of $\nu_e,\,\rho$ and $\theta$ is given by
\begin{equation} \label{eq:pdf_eigenvalues}
    p\left( \rho, \theta, \nu_{e} \right) = 225 \frac{\sqrt{5}}{2\pi} \rho \left( \rho^{2} - \theta^{2} \right)
    \exp \left[ - \frac{1}{2} \left( 15 \rho^{2} + 5 \theta^{2} + \nu_{e}^{2} \right) \right]
\end{equation}
after integrating over the irrelevant rotation angles that diagonalize the tidal tensor. Thus we can estimate the volume fractions taken up by each element of the cosmic web as:
\begin{equation} \label{eq:volume_fraction}
    F_{\mathrm{vol}} = \int_{0}^{\infty} d{\rho} \int_{-\rho}^{\rho} d{\theta}
    \int_{\nu_{\mathrm{th}} + f_{1}\left( \rho, \theta \given \alpha \right)}^{\nu_{\mathrm{th}} + f_{2}\left( \rho, \theta \given \alpha \right)}
    d{\nu_{e}} \; p(\rho, \theta, \nu_{e}).
\end{equation}

Now we use the Gaussian statistics to find the conditional Press-Schechter Ansatz.  The simplification that we are only interested in spherical collapse has the corollary that the distribution of the overdensity field conditional on environment only depends explicitly on the value of the environmental overdensity $\denv=\nu_e\sqrt{S_e}$:
\begin{equation} \label{eq:pdf_halo_env}
    p \left( \delta_{h} \given \denv \right) d{\delta_{h}} =
    \frac{d{\nu_{h}}}{\sqrt{2 \pi \left( 1 - \epsilon^{2} \right)}}
    \exp \left[ -\frac{ \left( \nu_{h} - \epsilon \nu_{e} \right)^{2} }{ 2 \left( 1 - \epsilon^{2} \right) } \right]
\end{equation}
where $\nu_{h} = \delta_{c}(S, \denv) / \sqrt{S}$ relates the halo mass $M$ to the corresponding value of the overdensity field $\delta_{h}$, with $\nu_{e} = \denv / \sqrt{\Senv}$ the analagous quantity for the environment and $\epsilon^{2} = S_{eh} / \left( \Senv S \right)$ is the correlation coefficient between $\delta_h$ and $\denv$.

We are now in a position to present the key integral of this paper.  First we must take into account that we require $S > \Senv$ for the excursion set theory to work.  This corresponds to the physical requirement that the halo be embedded in the environment and not vice-versa.  Under the simplification that the moving barrier $\delta_{c}$ does not affect \cref{eq:pdf_halo_env}, we find that the conditional collapsed mass fraction is that same as the result in \cref{eq:JAP_uncond_cmf}, subject to the change of variables from $\nu$ to $\nueff$ (defined below).  Thus we find the corollary that the first-crossing distribution becomes:
\begin{subequations} \label{eq:fcd_given_alpha}
\begin{align}
    g \left(S \given \alpha \right) &= \frac{1}{ F_{\mathrm{vol}} \left( \alpha \right) }
    \int_{0}^{\infty} d{\rho} \int_{-\rho}^{\rho} d{\theta}
    \int_{\nu_{\mathrm{th}} + f_{1}\left( \rho, \theta \given \alpha \right)}^{\nu_{\mathrm{th}} + f_{2}\left( \rho, \theta \given \alpha \right)} d{\nu_{e}} \;
    p(\rho, \theta, \nu_{e}) g \left( \nu_{\mathrm{eff}} \right)
\intertext{where, for the Peacock mass function:}
    g \left( \nueff \right) &= - \frac{\mathrm{exp} \left( c\nueff^{2} \right)}{(1 + a\nueff^{b})^{2}}  \left[ ab\nueff^{b-1} + 2c\nueff \left( 1+ a\nueff^{b} \right) \right]
      \frac{\partial{\nueff}}{\partial{S}}  \\
    \nu_{\mathrm{eff}} &= \mathrm{max} \left\{ 0, \frac{ \nu_{h} \left( S, \denv \right) - \epsilon^{2} \left( S, \Senv \right) \nu_{e}
    \left( \Senv, \denv \right) }{ \sqrt{ 1 - \epsilon^{2} \left( S, \Senv \right) }} \right\}
\end{align}
\end{subequations}
This is the first-crossing distribution which we substitute into \cref{eq:multiplicity} to obtain the $f(\R)$ multiplicity function conditional to the tidal classification of the environment.


\section{Simulations and algorithms} 
\label{sec:simulations_and_algorithms}

The simulations used in this paper were performed with the \texttt{ISIS} code \cite{2014A&A...562A..78L}, which is a modified gravity modification of the {\it N}-body code \texttt{RAMSES} \cite{2002A&A...385..337T}.

We performed two large-box simulations: one for \text{$\Lambda$CDM} and one for $\abs{f_{\R 0}} = 10^{-5}$ using an approximate method \cite{2015PhRvD..91l3507W} to incorporate the scalar field.  We also performed two small-box simulations: one for \text{$\Lambda$CDM} and one for $\abs{f_{\R 0}} = 10^{-5}$ fully solving for the scalar field in the simulation box \cite{2015MNRAS.454.4208W}. The approximate method has previously been shown to produce very good results (accuracy to a few percent) with respect to the matter power spectrum and the halo mass function \cite{2015PhRvD..91l3507W}. The modified gravity simulations were performed using the same inital conditions as the corresponding \text{$\Lambda$CDM} simulations. This allows us to compare the two models without requiring multiple realisations of the intial conditions.  The cosmological parameters are:  $\Omega_{m0} = 0.27$, $\Omega_{\Lambda 0} = 0.73$, $h = 0.704$, $n_{s} = 0.966$ and $\sigma_8 = 0.8$ and the other simulation parameters are listed in \cref{tab:simulation_params}.

We located haloes using a Friend-of-Friend algorithm\footnote{\url{https://github.com/damonge/MatchMaker}} as well as the spherical overdensity halo finder \texttt{AHF} (Amiga Halo Finder)\footnote{\url{http://popia.ft.uam.es/AHF/Download.html}} \cite{2004MNRAS.351..399G}. The {\it N}-body particles were binned to a $N_{\mathrm{grid}} = 512^{3}$ grid using Cloud-in-Cell interpolation and then smoothed with a Gaussian kernel of widths $R_{\rm env} = 10$ and $20$ Mpc$/h$ in this paper. The local tidal tensor and its eigenvalues were then computed using a public code\footnote{\url{https://github.com/damonge/DensTools}}. The error bars in the plots were computed using jack-knife resampling of the simulation box with 8 samples. To generate error-bars for ratio plots coming from two different simulations, we used:
\begin{equation} \label{eq:uncorrelated_errors}
    \Delta \left( \frac{a}{b} \right) = \frac{a}{b} \sqrt{\left( \frac{\Delta a}{a} \right)^{2} + \left( \frac{\Delta b}{b} \right)^{2} - 2\,r\, \frac{\Delta a}{a} \frac{\Delta b}{b}}
\end{equation}
where $r$ is the correlation coefficient. The quantities whose ratios we compute are highly correlated (since they are taken from simulations with the same initial conditions) so for simplicy we have simply used a fixed value of $r = 0.9$ (or $r = 0$ for cases where we wanted to be conservative) to estimate the error-bars.

\begin{table}
\begin{tabular}{l l l l}
\toprule
Box size (Mpc/h) & Particles per box & Cosmology & MG algorithm \\
\midrule
$1024$ & $1024^{3}$ & \text{$\Lambda$CDM}, $\abs{f_{\R 0}} = 10^{-5}$ & Approximate \\
$256$ & $512^{3}$ & \text{$\Lambda$CDM}, $\abs{f_{\R 0}} = 10^{-5}$ & Full \\
\bottomrule \\[-.1in]
\end{tabular}
\caption{The {\it N}-body simulations used in this paper. The cosmological parameters used to generate initial conditions was:  $\Omega_{m0} = 0.27$, $\Omega_{\Lambda 0} = 0.73$, $h = 0.704$, $n_{s} = 0.966$ and $\sigma_8 = 0.8$.  \label{tab:simulation_params}}
\end{table}


\section{Results} 
\label{sec:results}

We compare the behaviour of the \text{$\Lambda$CDM} and $f(\R)$ mass functions in various environments.  Then we evaluate the accuracy of our semi-analytic model using {\it N}-body simulations and describe its limitations and possibles paths to improvement. Finally, we analyze the effect of modified gravity on a number of internal halo properties as a function of tidal environment.

\subsection{Volume fractions} 
\label{sub:volume_fractions}

\begin{figure}[]
    \begin{center}
        \subfloat[$\Renv = 20/h$ Mpc]{\label{subfig:plot_volfrac_all_renv_20}\includegraphics[keepaspectratio,width=0.7\textwidth]{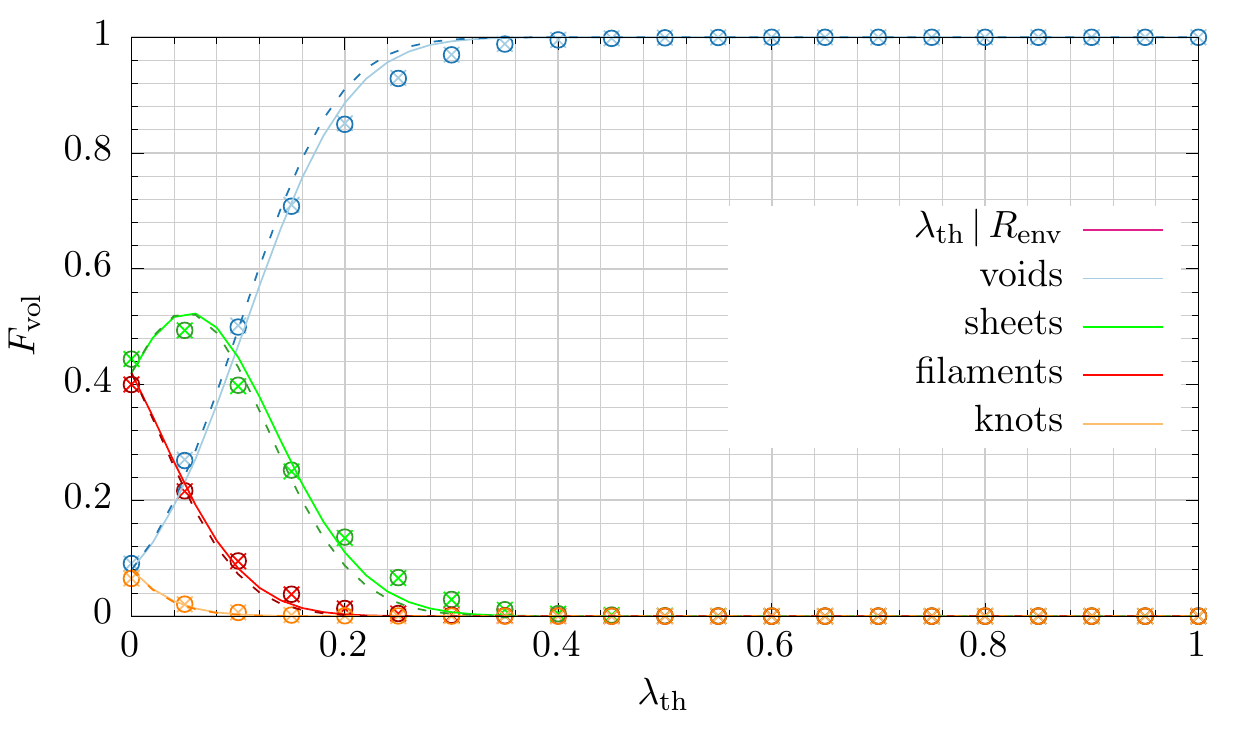}} \\
        \subfloat[$\Renv = 10/h$ Mpc]{\label{subfig:plot_volfrac_all_renv_10}\includegraphics[keepaspectratio,width=0.7\textwidth]{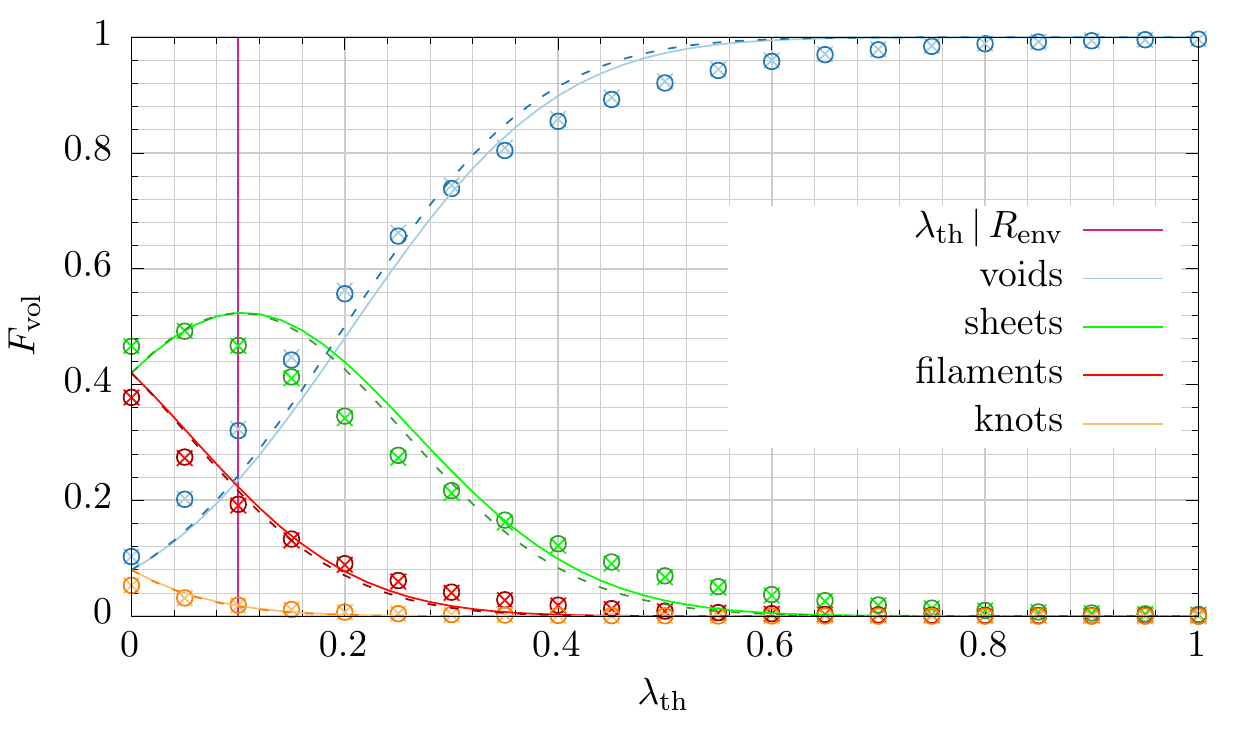}}
    \end{center}
    \caption{Volume fraction according to the Gaussian probability distribution function \cref{eq:pdf_eigenvalues} (lines) and the {\it N}-body simulations (points).  Light colours (solid lines; crosses) are for \lcdm{}, darker ones (dashed lines; circles) for $\fr[5]$.}
    \label{fig:comp_volfrac_all}
\end{figure}

We can indirectly measure the accuracy of the Gaussian model for the eigenvalue probability distribution via the volume fractions in different cosmic web elements.  In \cref{fig:comp_volfrac_all} we compare the values from our model to those from the {\it N}-body simulations.

The behaviour with $\lambda_{\mathrm{th}}$ permits us to select a suitable value for $\nu_{\mathrm{th}}$ in \cref{eq:nueff_integral_limits}.  We require a value for which there is a (relatively) even distribution between the different morphologies (amongst other requirements discussed in \cite{Alonso:2014zfa}).  Thus we select $\lambda_{\mathrm{th}} = 0$ for $\Renv = 20$ Mpc$/h$ and $\lambda_{\mathrm{th}} = 0.1$ for $\Renv = 10$ Mpc$/h$.  We utilise the same threshold for the smoothing in our semi-analytic model and the {\it N}-body simulations.

We compare the volume fractions calculated from \cref{eq:volume_fraction} to the {\it N}-body volume fraction in \cref{fig:comp_volfrac_all}. At scales of $R_{\rm env}=20\,{\rm Mpc}/h$ \cref{eq:pdf_eigenvalues} performs well, whereas at $R_{\rm env}=10\,{\rm Mpc}/h$ we find that $p(\rho, \theta, \phi)$ is a poor approximation of the real overdensity field.  This should not be surprising given the non-linear evolution of the density field on all but the largest scales.

The Gaussian volume fraction in \cref{fig:comp_volfrac_all} illustrates the large-scale behaviour of the scalar field modification to general relativity.  Compared to \text{$\Lambda$CDM} (solid lines in \cref{fig:comp_volfrac_all}) there is a greater fraction of voids and a smaller fraction in collapsed structures.  At larger scales, the change in volume fraction due to modified gravity is more pronounced.  This is due to the scalar field leaving the chameleon regime at small scales and moving to the linear regime at large scales (although on cosmological scales it returns to mimic general relativity behaviour).  More compact morphologies exhibit a greater deficiency, so that the overall fraction remains at unity.  However, the difference between even $\abs{f_{\R 0}} = 10^{-5}$ and \text{$\Lambda$CDM} is dwarfed by the difference between the theoretical volume fraction and the {\it N}-body result at all smoothing scales.  This provides a quantitative basis for the assertion in \cite{Lombriser:2013wta} that there is insufficient evidence to warrant modelling the environemnt density using the $f(\R)$ equations.


\subsection{Multiplicity functions in each environment} 
\label{sub:multiplicity_functions_in_each_environment}

\begin{figure}
    \begin{center}
        \includegraphics[keepaspectratio,width=0.75\textwidth]{{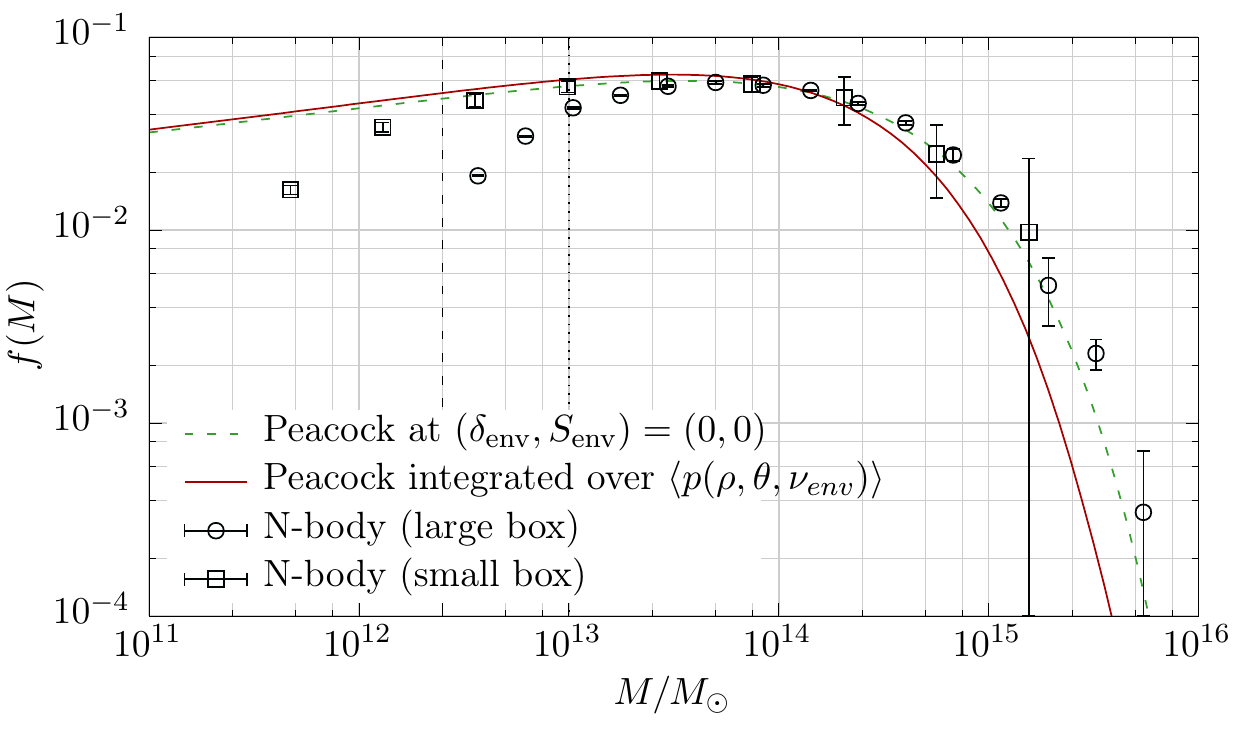}}
    \end{center}
    \caption{Multiplicity function $f(M)$ for each \lq\lq{}unconditional\rq\rq{} definition at $R_{\rm env}=10\,{\rm Mpc}/h$. The green dashed line shows the Peacock model for the unconditional multiplicity function, while the solid red line shows our re-scaling method for the conditional mass function applied to the Peacock multiplicity function and then integrated over the environment distribution. Boxes and circles show the data from the large- and small-box {\it N}-body simulations. The cut-off below which the mass resolution of the simulations make the results unreliable are shown as dashed (dotted) lines for the small (big) simulations.}
    \label{fig:dfdlnm_univ_20}
\end{figure}

\begin{figure}
    \begin{center}
        \subfloat[\lcdm{}]{ 
            \label{subfig:comp_dfdlnm_lcdm_R_20}
            \includegraphics[keepaspectratio,width=0.49\textwidth]{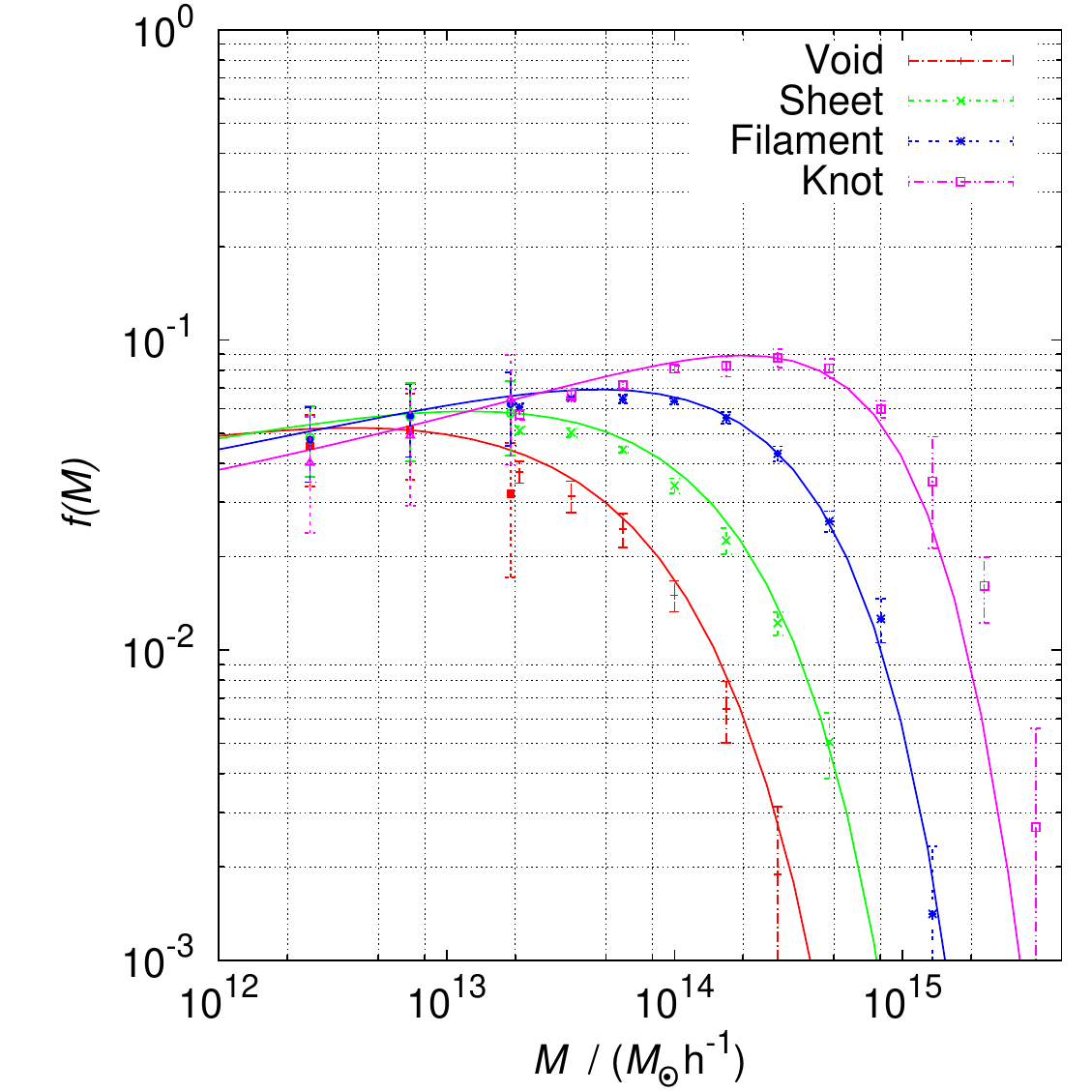}
        }
        \subfloat[{$\fr[5]$}]{ 
            \label{subfig:comp_dfdlnm_f5_R_20}
            \includegraphics[keepaspectratio,width=0.49\textwidth]{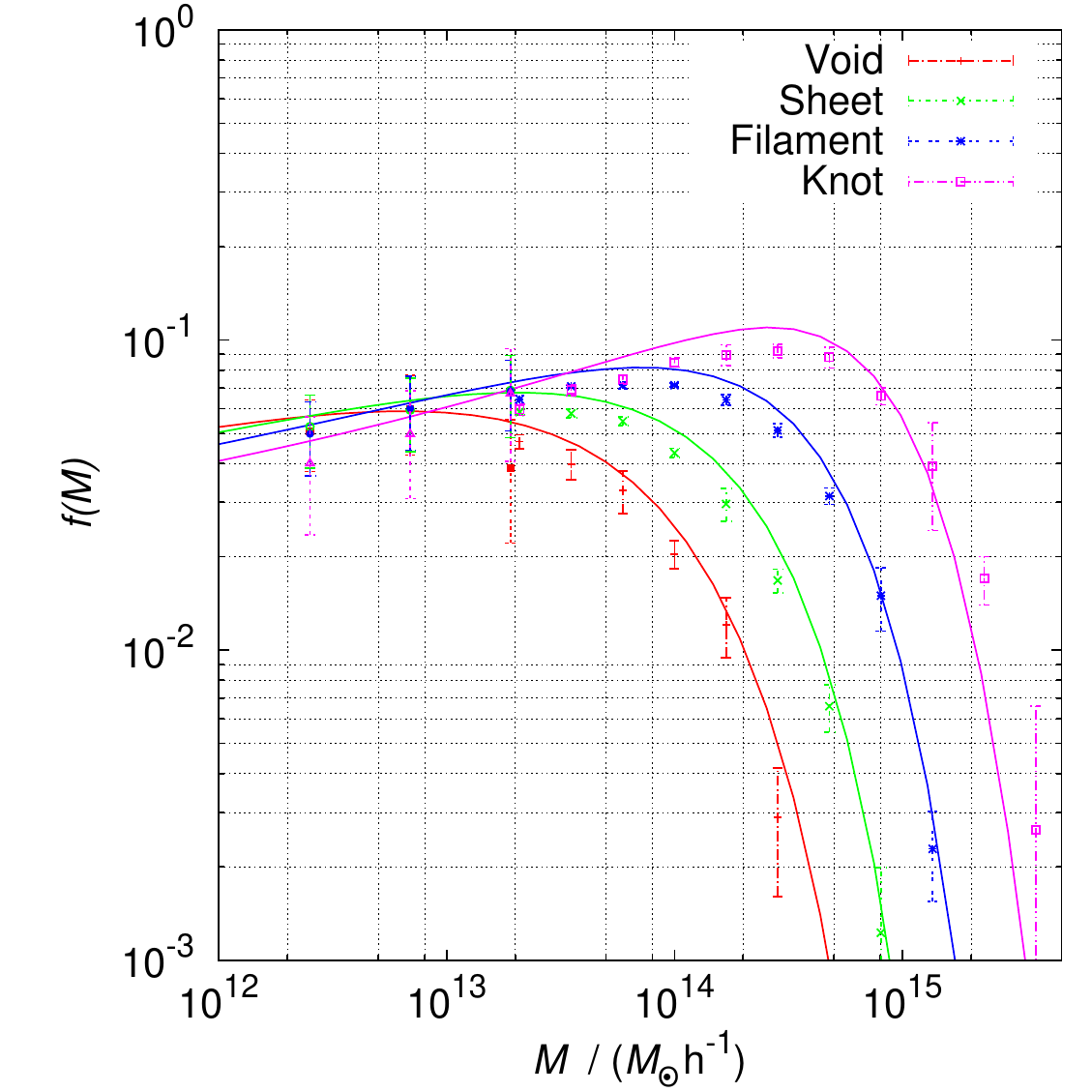}
        } \\
        \subfloat[{$\fr[5]/$\lcdm{}}]{ 
            \label{subfig:comp_dfdlnm_ratio_f5_R_20}
            \includegraphics[keepaspectratio,width=0.49\textwidth]{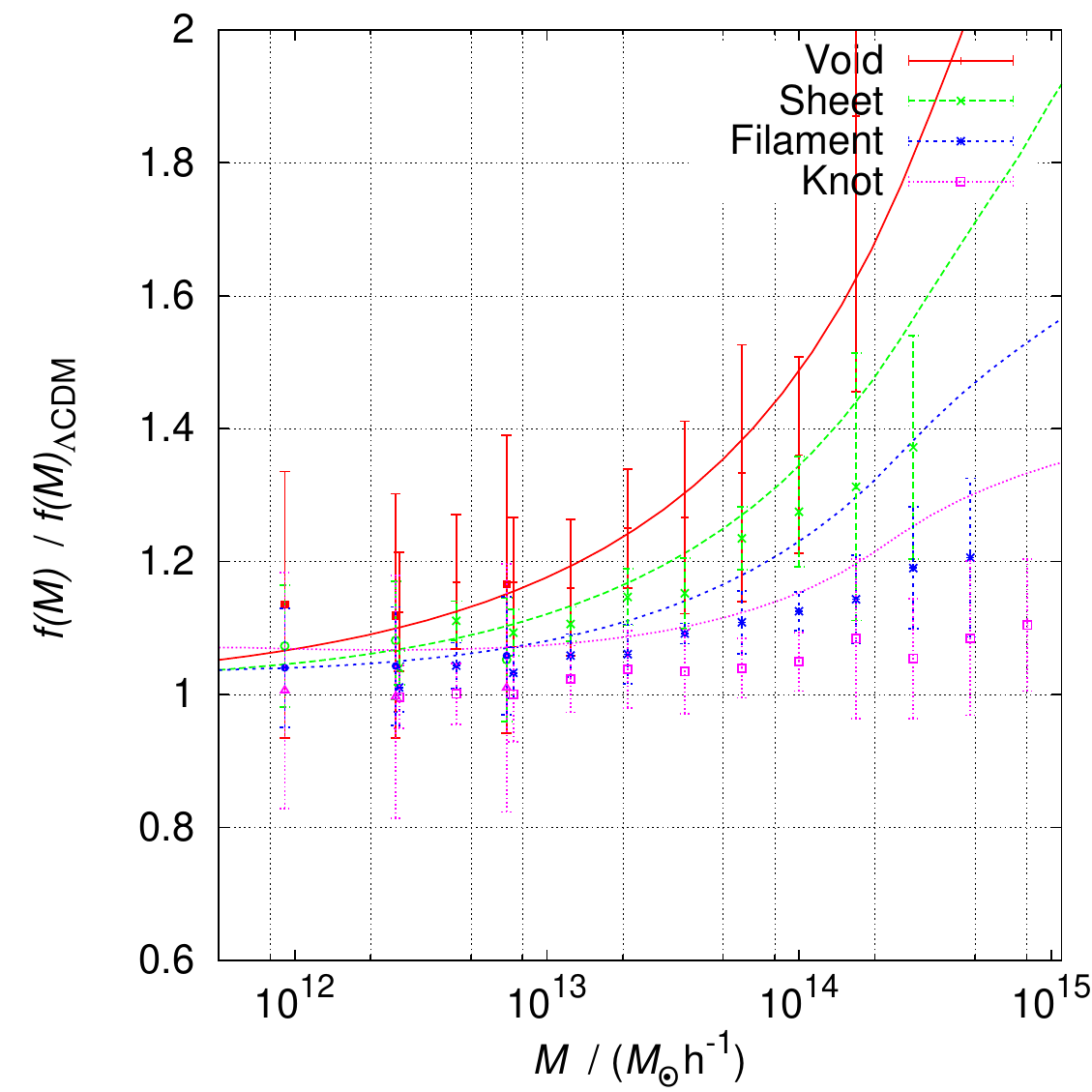}
        }
        \end{center}
    \caption{$f(M)$ for each environment at $R_{\rm env}=20\,{\rm Mpc}/h$, assuming a Gaussian window function.  Lines indicate our model for \text{$\Lambda$CDM} (left) and $\abs{f_{\R 0}} = 10^{-5}$ (right) according to \cref{eq:fcd_given_alpha} and points indicate the {\it N}-body results.}
    \label{fig:dfdlnm_renv_20}
\end{figure}

\begin{figure}
    \begin{center}
        \subfloat[\lcdm{}]{ 
            \label{subfig:comp_dfdlnm_lcdm_R_20_sharpk}
            \includegraphics[keepaspectratio,width=0.49\textwidth]{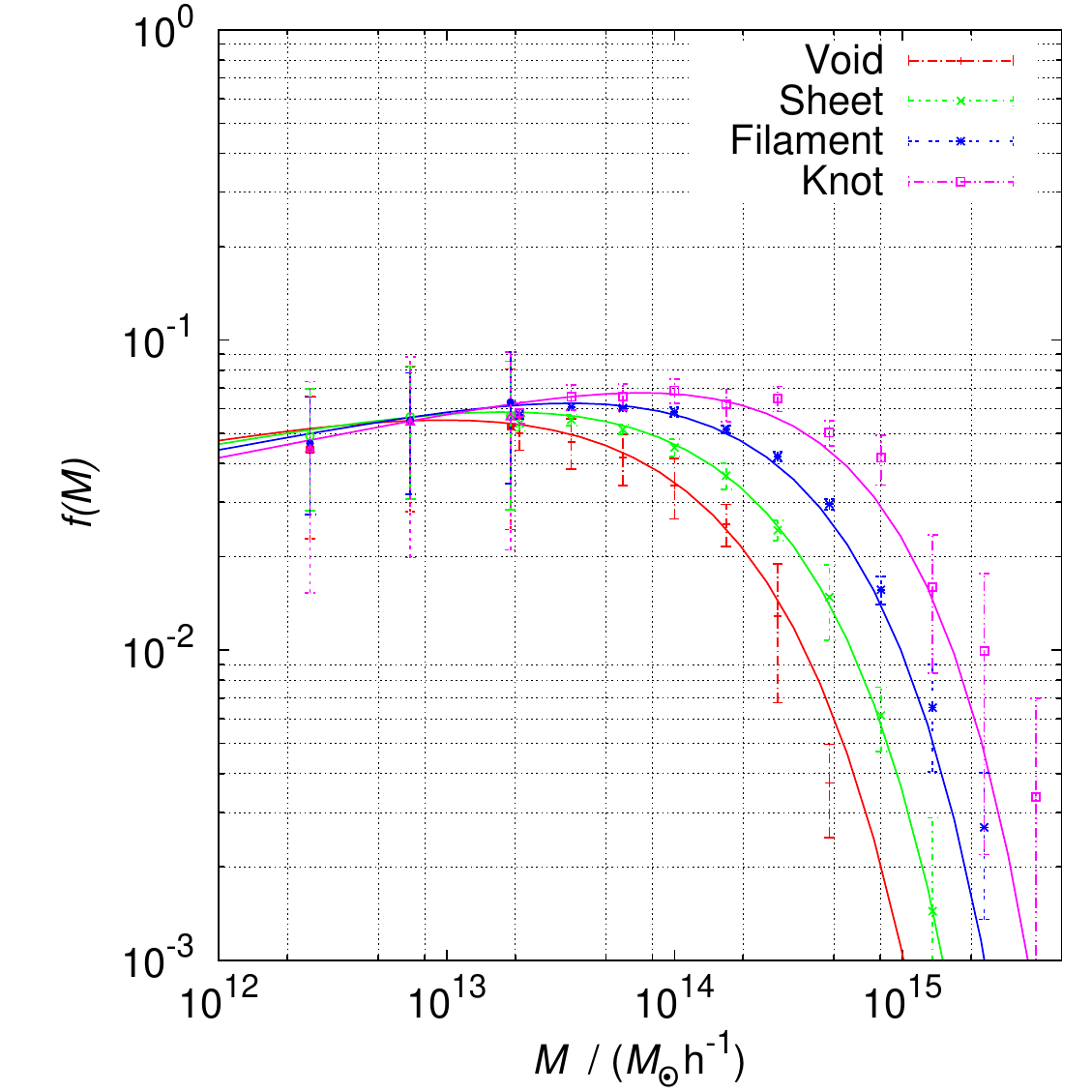}
        }
        \subfloat[{$\fr[5]$}]{ 
            \label{subfig:comp_dfdlnm_f5_R_20_sharpk}
            \includegraphics[keepaspectratio,width=0.49\textwidth]{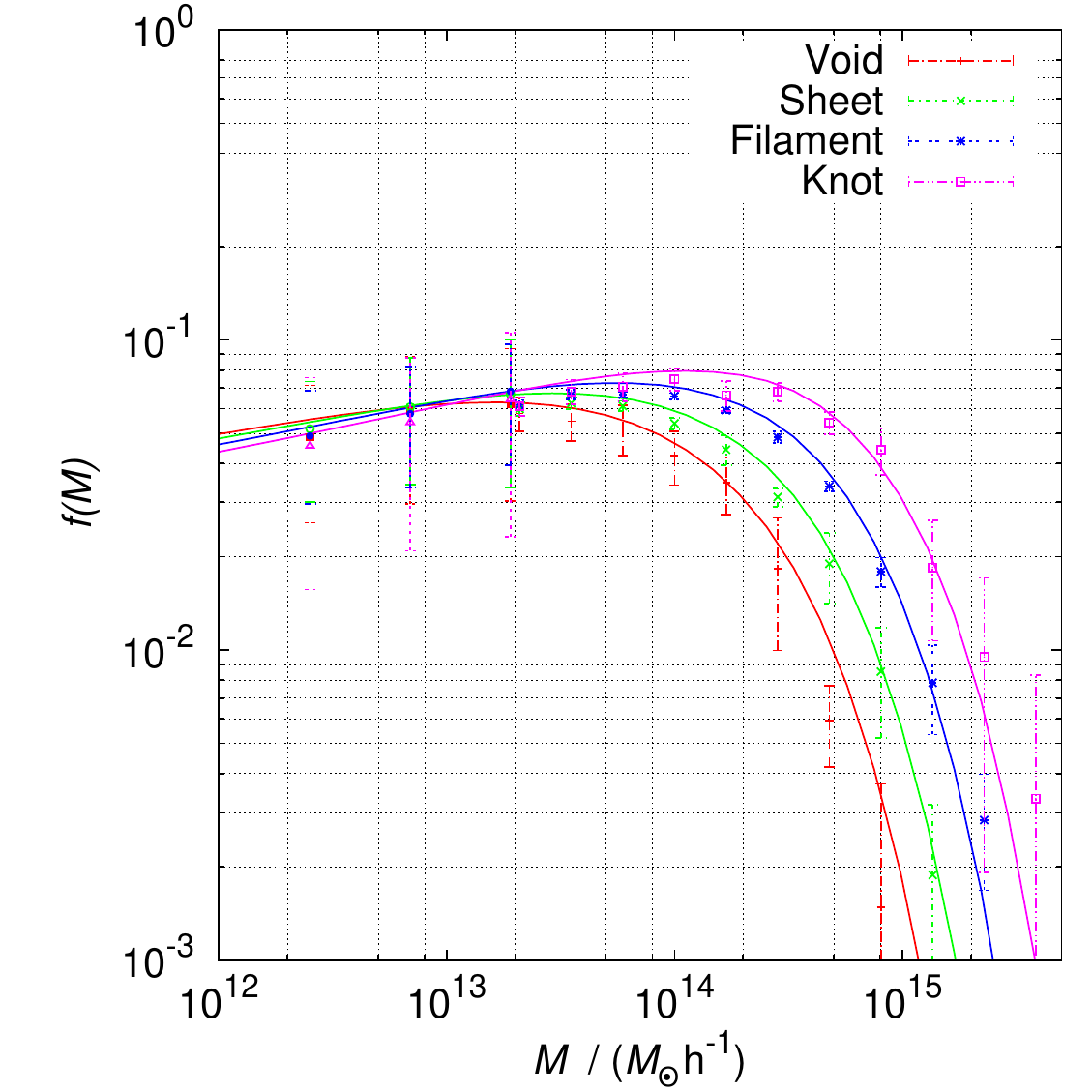}
        } \\
        \subfloat[{$\fr[5]/$\lcdm{}}]{ 
            \label{subfig:comp_dfdlnm_ratio_f5_R_20_sharpk}
            \includegraphics[keepaspectratio,width=0.49\textwidth]{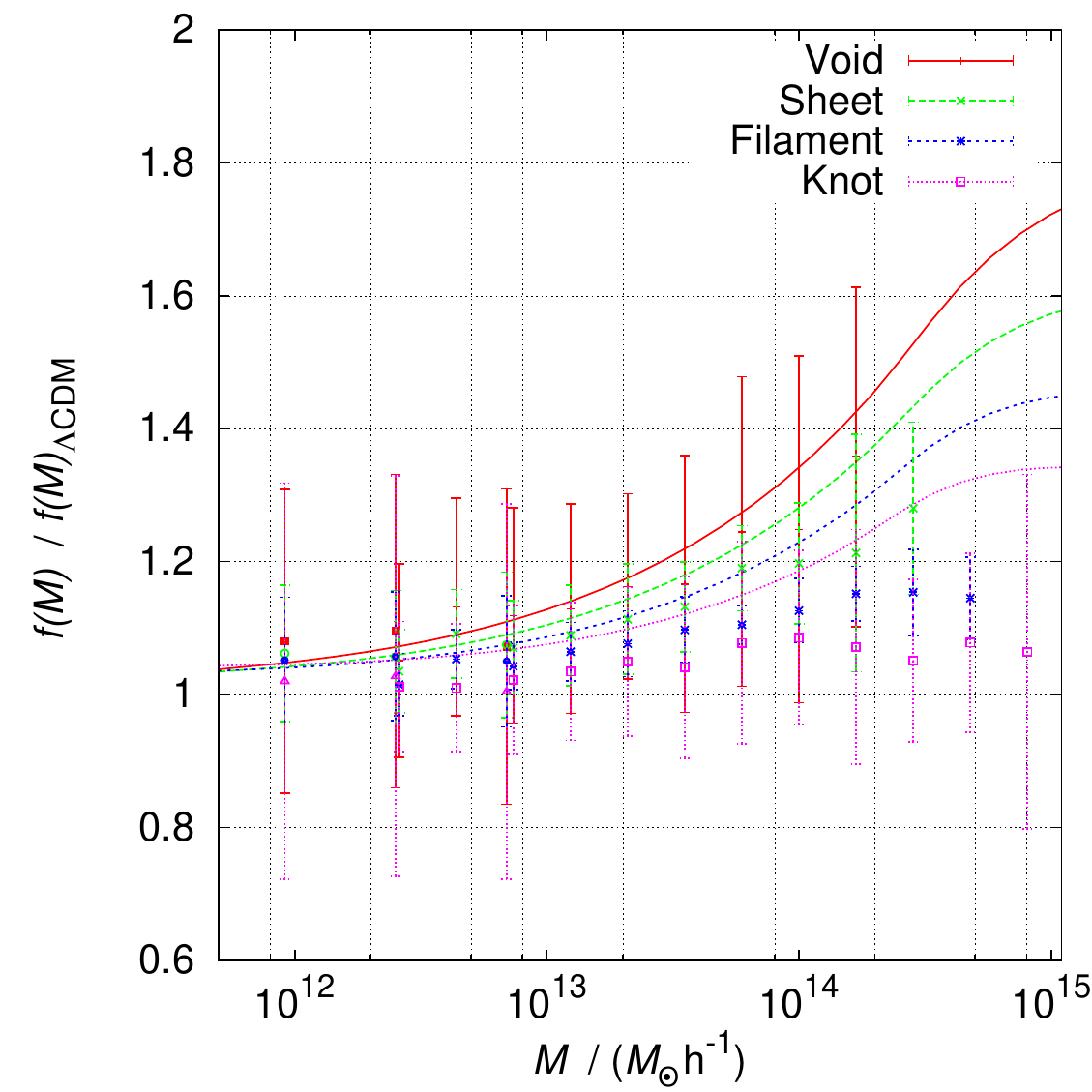}
        }
        \end{center}
    \caption{$f(M)$ for each environment at $R_{\rm env}=20\,{\rm Mpc}/h$, assuming a sharp-$k$ window function.  Lines indicate our model for \text{$\Lambda$CDM} (left) and $\abs{f_{\R 0}} = 10^{-5}$ (right) according to \cref{eq:fcd_given_alpha} and points indicate the {\it N}-body results.}
    \label{fig:dfdlnm_renv_20_sharpk}
\end{figure}

\begin{figure}
    \begin{center}
        \subfloat[\lcdm{}]{ 
            \label{subfig:comp_dfdlnm_lcdm_R_10}
            \includegraphics[keepaspectratio,width=0.49\textwidth]{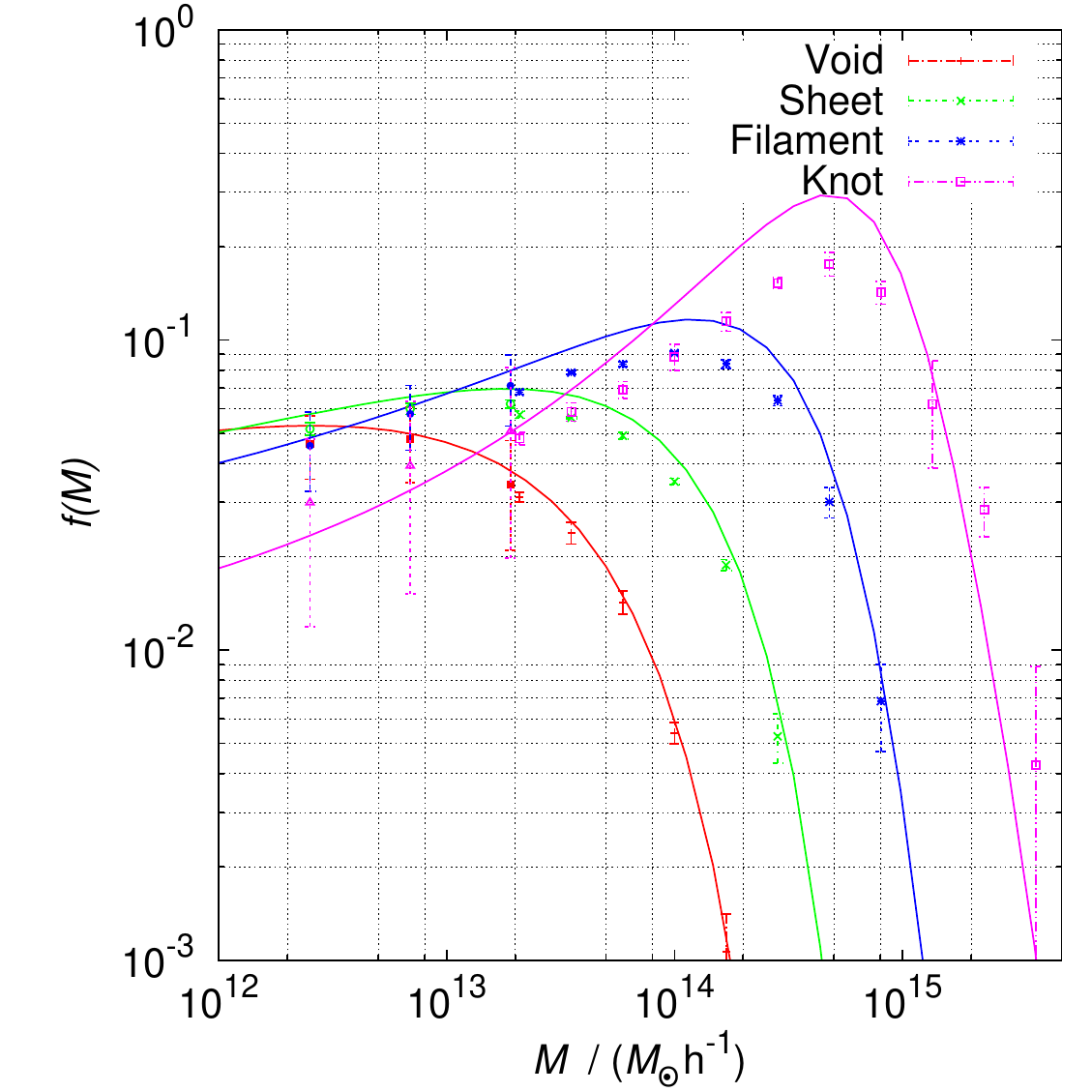}
        }
        \subfloat[{$\fr[5]$}]{ 
            \label{subfig:comp_dfdlnm_f5_R_10}
            \includegraphics[keepaspectratio,width=0.49\textwidth]{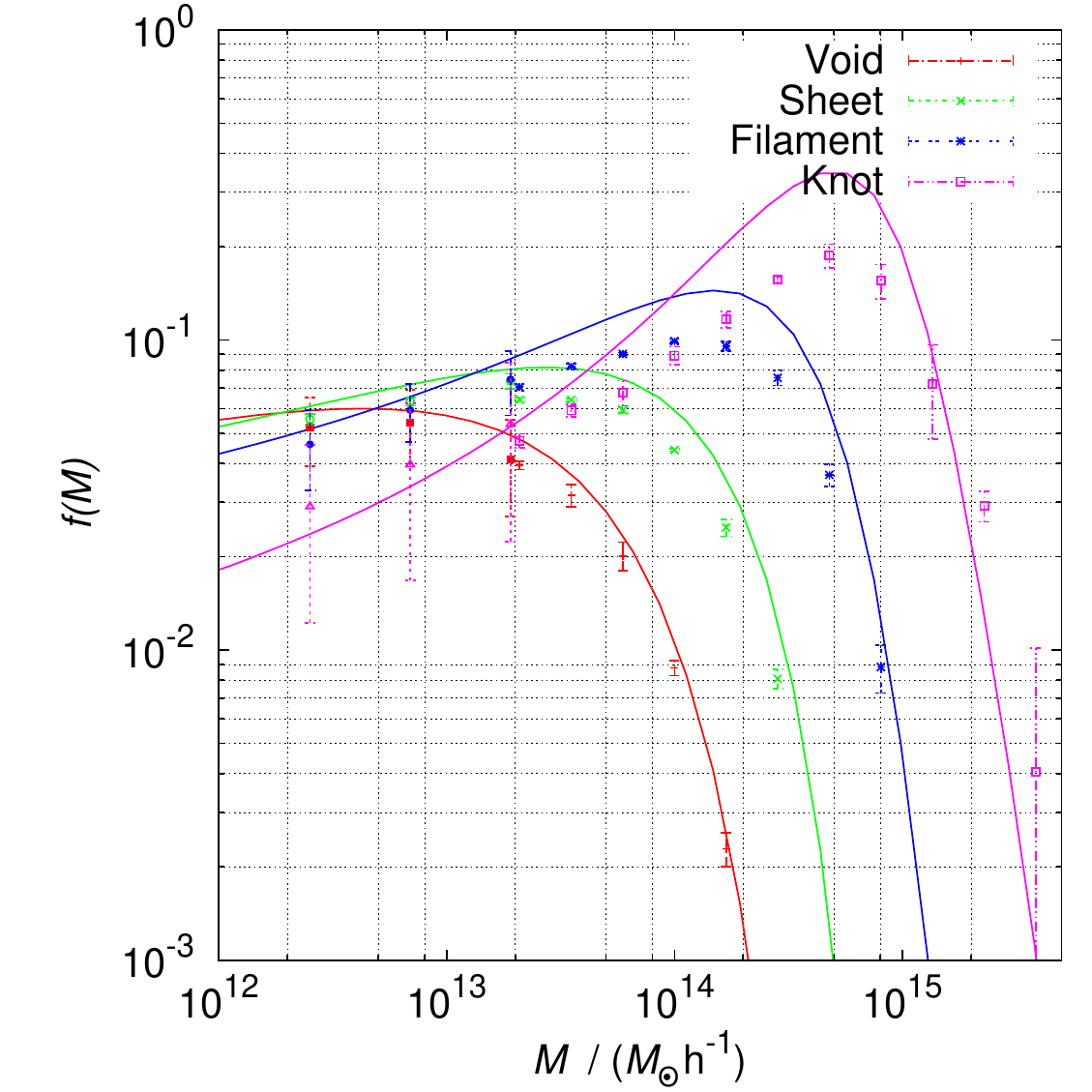}
        } \\
        \subfloat[{$\fr[5]/$\lcdm{}}]{ 
            \label{subfig:comp_dfdlnm_ratio_f5_R_10}
            \includegraphics[keepaspectratio,width=0.49\textwidth]{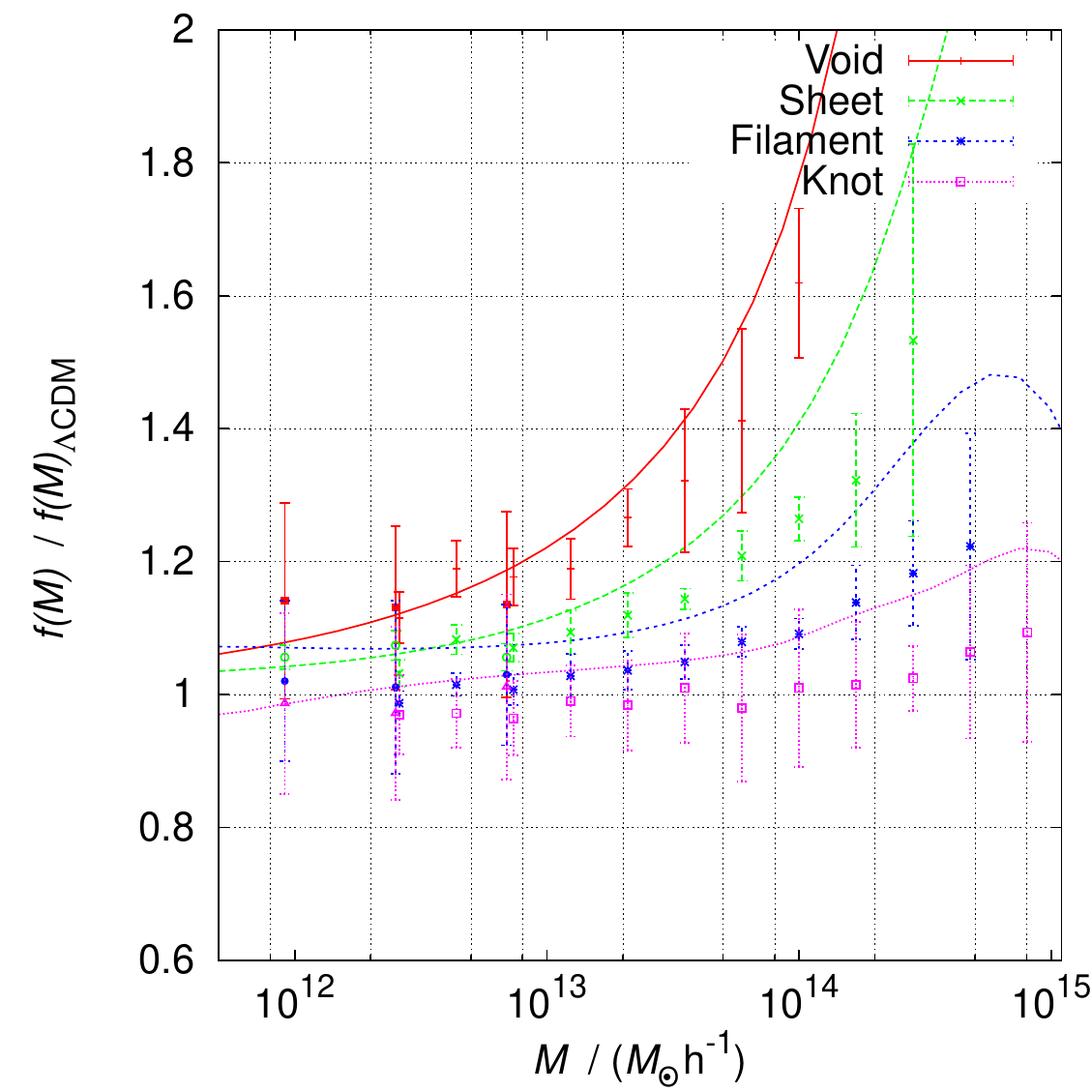}
        }
        \end{center}
    \caption{$f(M)$ for each environment at $R_{\rm env}=10\,{\rm Mpc}/h$, assuming a Gaussian window function.  Lines indicate our model for \text{$\Lambda$CDM} (left) and $\abs{f_{\R 0}} = 10^{-5}$ (right) according to \cref{eq:fcd_given_alpha} and points indicate the {\it N}-body results.}
    \label{fig:dfdlnm_renv_10}
\end{figure}

\begin{figure}
    \begin{center}
        \subfloat[\lcdm{}]{ 
            \label{subfig:comp_dfdlnm_lcdm_R_10_sharpk}
            \includegraphics[keepaspectratio,width=0.49\textwidth]{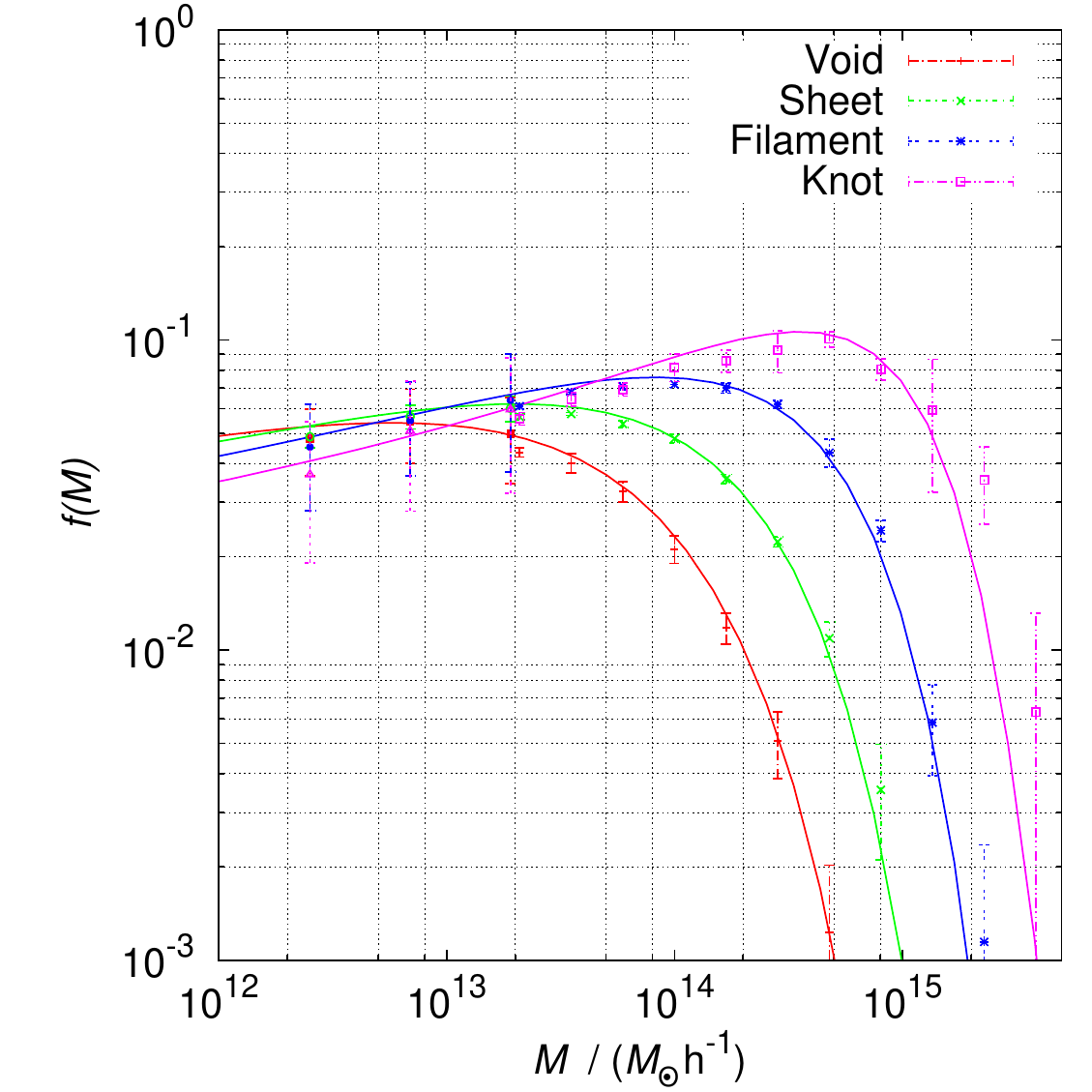}
        }
        \subfloat[{$\fr[5]$}]{ 
            \label{subfig:comp_dfdlnm_f5_R_10_sharpk}
            \includegraphics[keepaspectratio,width=0.49\textwidth]{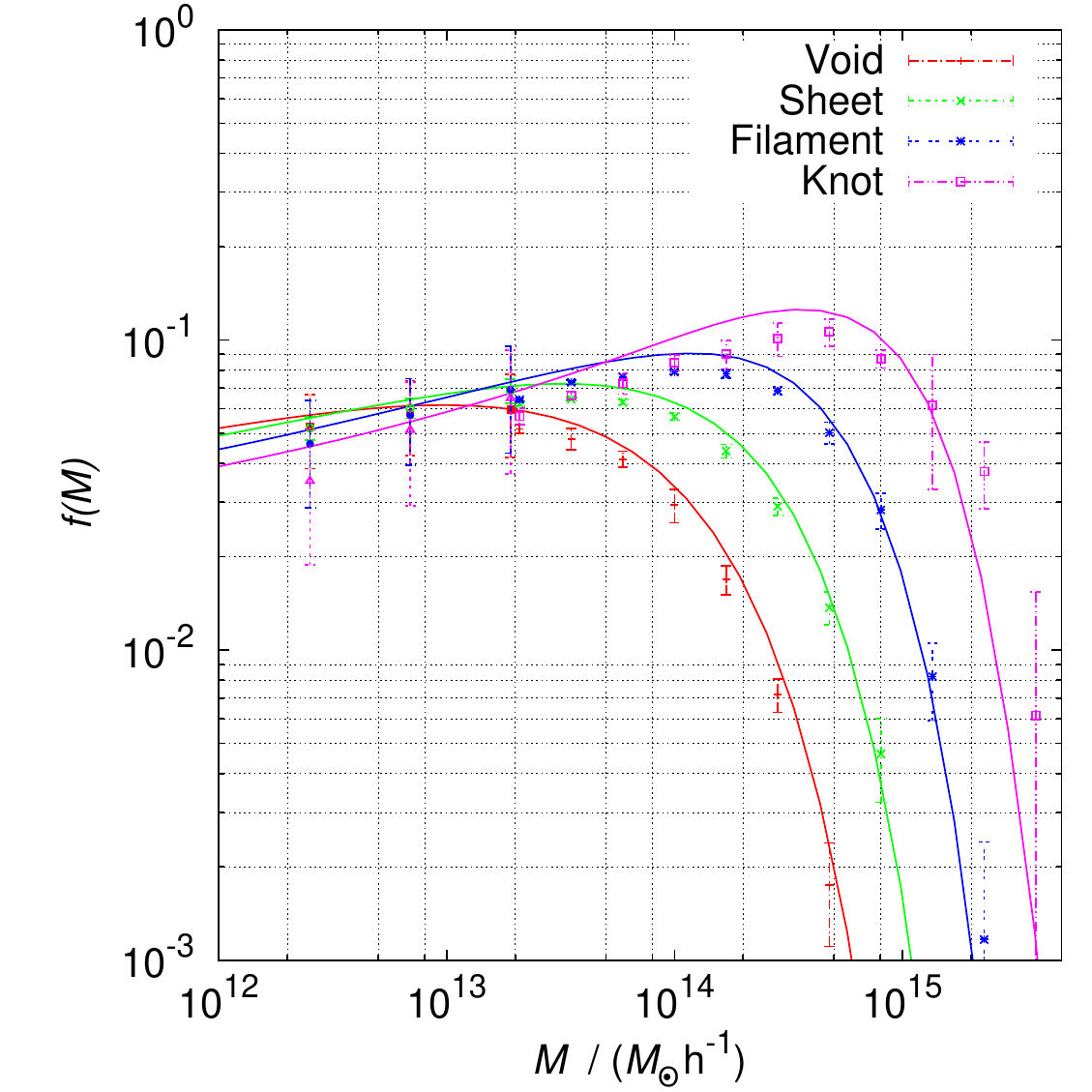}
        } \\
        \subfloat[{$\fr[5]/$\lcdm{}}]{ 
            \label{subfig:comp_dfdlnm_ratio_f5_R_10_sharpk}
            \includegraphics[keepaspectratio,width=0.49\textwidth]{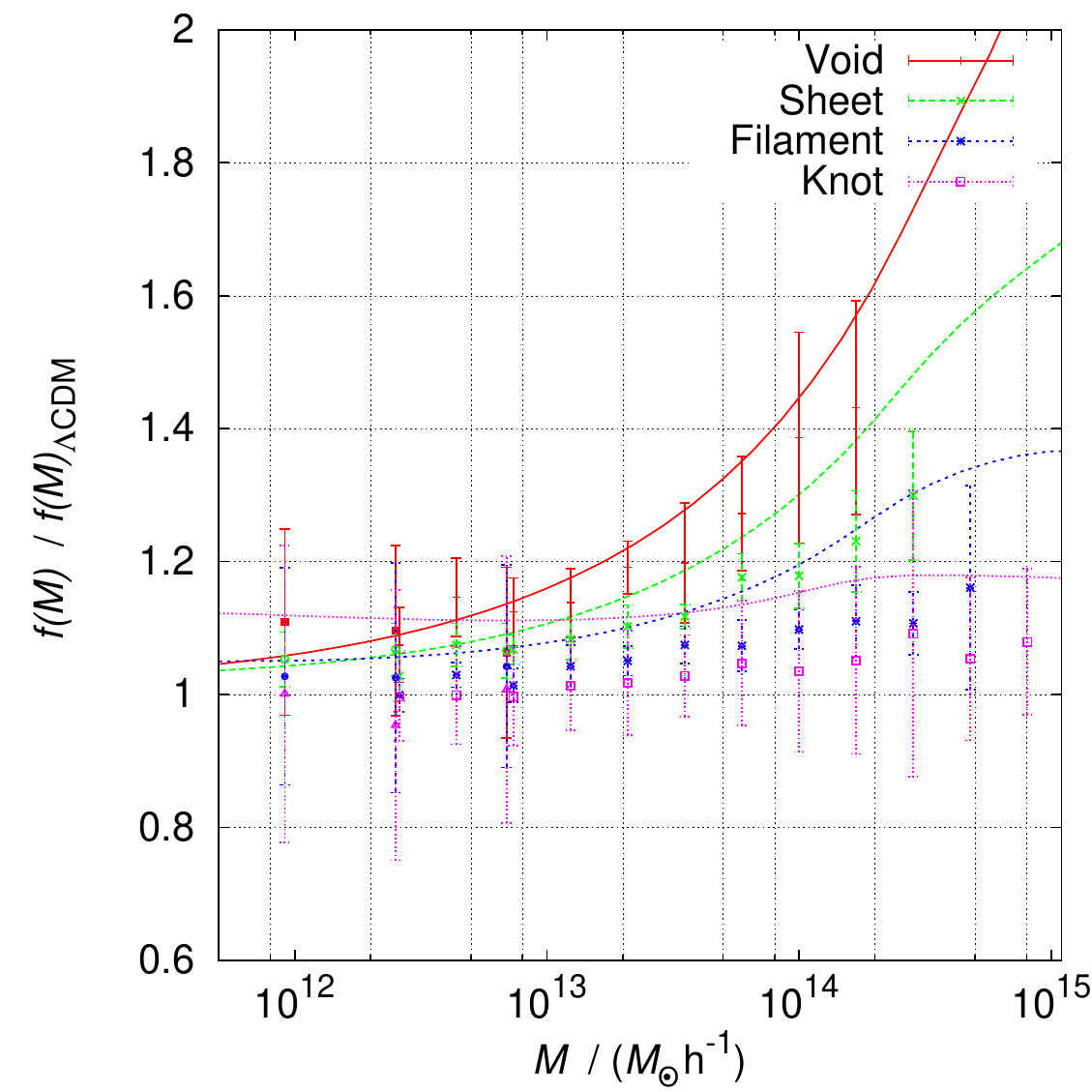}
        }
        \end{center}
    \caption{$f(M)$ for each environment at $\Renv[10]$, assuming a sharp-$k$ window function..  Lines indicate our model for \lcdm{} (solid) and $\fr[5]$ (dashed) according to \cref{eq:fcd_given_alpha} and points indicate the {\it N}-body results.}
    \label{fig:dfdlnm_renv_10_sharpk}
\end{figure}

We compare the behaviour of the \text{$\Lambda$CDM} mass functions to $f(\R)$ with $\abs{f_{\R 0}} = 10^{-5}$ in the different structures of the cosmic web and averaged over all environments.  Our semi-analytic model performs best at successively larger scales and in lower-density environments, but reproduces the main trend found in the simulations: the $f(\R)$ abundances are amplified in underdense environments and at low masses.

The smoothing scale determines the range of masses in which our results are reliable. Recall from \cref{sub:the_excursion_set_theory_in_lcdm} that the major shortcoming of the excursion set formalism is that whenever $S < \Senv$ the process is not well defined, since the condition $S < \Senv$ implies that the halo has already collapsed at resolution $S$, before the starting points of its trajectory at $\Senv$. Therefore care must be taken when interpreting any results for masses with $S\sim\Senv$. For the filter sizes used in this work, $R=10\,{\rm Mpc}/h$ and $R=20\,{\rm Mpc}/h$, the corresponding masses enclosed within them are $M_{10}=1.67\cdot 10^{15}\,M_\odot/h$ and $M_{20}=1.34\cdot 10^{16}\,M_\odot/h$, which act as upper bounds for the mass range where the excursion set predictions are meaningful.

A desirable property of any model of the mass function that incorporates environmental dependence is that, when marginalised over all possible environments, one should recover the unconditional mass function given by the same model. Currently our model is based on simply substituting the argument of the unconditional mass function from $\nu_{h}$ to $\nu_{\rm eff}$, inspired by the equivalent result found in the excursion-set formalism. However, this rescaling of the mass function is only mathematically consistent, in the sense described above, for the specific form of the Press-Schechter mass function \cite{2011PhRvD..83f3511P}. Thus, most simple attempts at modelling the conditional mass function have only been able to provide a qualitative description of it, with a poor quantitative performance. This problem is also illustrated in Figure \ref{fig:dfdlnm_univ_20}, which shows the mass function from our simulations together with the Peacock unconditional mass function and the excursion set prediction after rescaling the Peacock mass function and marginalising it over the environment. Our results agree with \cite{2011PhRvD..83f3511P}, who showed that the only solution of \cref{eq:volterra_fcd} which matches the unconditional case is Press-Schechter with a $\delta_{c}$ linear in $S$.  Thus we do not expect the cosmic-web-sensitve result to match the percent-level fits to N-body which motivated the unconditional form of the Peacock mass function.

The additional dependence of the collapse density in $f(\R)$ amplifies this same problem: compare \cref{subfig:comp_dfdlnm_lcdm_R_20,subfig:comp_dfdlnm_lcdm_R_10,subfig:comp_dfdlnm_lcdm_R_20_sharpk,subfig:comp_dfdlnm_lcdm_R_10_sharpk} to \cref{subfig:comp_dfdlnm_f5_R_20,subfig:comp_dfdlnm_f5_R_10,subfig:comp_dfdlnm_f5_R_20_sharpk,subfig:comp_dfdlnm_f5_R_10_sharpk}, which show the multiplicity functions in different environments measured from the simulations together with the predictions of our method for a range of smoothing kernels (with $R_{\rm env}=10$ and 20 ${\rm Mpc}/h$ in \cref{fig:dfdlnm_renv_10,fig:dfdlnm_renv_10_sharpk} and \cref{fig:dfdlnm_renv_20,fig:dfdlnm_renv_20_sharpk} respectively, and using Gaussian and sharp-$k$ filters in \cref{fig:dfdlnm_renv_10,fig:dfdlnm_renv_20} and \cref{fig:dfdlnm_renv_10_sharpk,fig:dfdlnm_renv_20_sharpk} respectively). We concur with \cite{Lombriser:2013wta}, who state that the environment-averaged result simplifies to the result with the random walk starting at $\left( \denv = 0, \Senv = 0 \right)$ (i.e. to the unconditional mass function) only in $\Lambda$CDM and not in $f(\R)$.  Thus we expect our fits to be less accurate (when compared to {\it N}-body results) in $f(\R)$ than in $\Lambda$CDM, a hypothesis borne out by our results.

The trends with environment smoothing scale and density shown in \cref{fig:dfdlnm_renv_20,fig:dfdlnm_renv_20_sharpk,fig:dfdlnm_renv_10,fig:dfdlnm_renv_10_sharpk} can be understood in terms of deviations of the non-linear environment distribution with respect to the Gaussian prediction. Having seen in \cref{fig:comp_volfrac_all} that the $R_{\rm env}=20\,{\rm Mpc}/h$ volume fraction is well-approximated by the integral of $p(\rho, \theta, \nu_{e})$, the density field on such large scales in still evolving linearly from the primordial Gaussian density field.
Thus in \cref{fig:dfdlnm_renv_20,fig:dfdlnm_renv_20_sharpk}, we find good agreement in both \text{$\Lambda$CDM} and $f(\R)$, regardless of halo mass or the location within the cosmic web.  On environment scales of $R_{\rm env}=10\,{\rm Mpc}/h$, the results in \cref{sub:volume_fractions} suggest that the density field evolution is weakly-linear.  We expect deviations from linear collapse to occur more quickly in increasingly dense environments, where the local structures have already collapsed along one (sheets), two (filaments) or all three (knots) axes.  In \cref{fig:dfdlnm_renv_10,fig:dfdlnm_renv_10_sharpk} our predictions lie within the {\it N}-body error bars at progressively fewer points for voids, sheets, filaments and knots.  These trends occur independent of cosmology, which futher reinforces the hypothesis that the Gaussian model for the density is responsible, as we model the large-scale structure according to general relativity in both \text{$\Lambda$CDM} and modified gravity.

Like \cite{Alonso:2014zfa}, we find that the excursion-set prediction for the conditional mass function in \text{$\Lambda$CDM} becomes a worse model for the {\it N}-body data in higher-density environments, underpredicting the number of low-mass halos and over-predicting the abundance in the high-mass tail at $R_{\rm env}=10\,{\rm Mpc}/h$ (see \cref{subfig:comp_dfdlnm_lcdm_R_10,subfig:comp_dfdlnm_f5_R_10}), and this is true also for $f(\R)$.  For a smoothing of $20\,{\rm Mpc}/h$, \cref{subfig:comp_dfdlnm_lcdm_R_20,subfig:comp_dfdlnm_f5_R_20} show that the prediction becomes a much better fit, with a slight deficiency of haloes at high masses in knots.

Despite the shortcomings of the excursion set formalism when making accurate predictions for the conditional mass function, we can still use this framework to understand the differences between the $\Lambda$CDM and $f(\R)$ predictions. The $f(\R)$ mass functions are amplified compared to \text{$\Lambda$CDM}.  The key trends visible in \cref{subfig:comp_dfdlnm_ratio_f5_R_20,subfig:comp_dfdlnm_ratio_f5_R_10,subfig:comp_dfdlnm_ratio_f5_R_10_sharpk,subfig:comp_dfdlnm_ratio_f5_R_20_sharpk} are that the amplification increases with lower environment density. The amplification tends to increase with halo mass in voids and sheets, while there is a maximum enhancement mass in filaments and knots. This behaviour can be explained by the peak-background split.  For equivalent values of $S$, the barrier density is lower for $\abs{f_{\R 0}} = 10^{-5}$ than \text{$\Lambda$CDM}, which makes it more probable for the first-crossing of the excursion set trajectory to occur.  This effect is amplified by a larger value of $\delta_{c}(\denv, S) - \denv$, \textit{i.e.} in voids and sheets, but damped in dense regions where $\delta_{c}(\denv, S) - \denv \rightarrow 0$ via screening. Thus, the modified gravity behaviour compared to \text{$\Lambda$CDM} is driven by screening at all scales.

In order to improve the semi-analytical model for the mass function used here, three problems need to be overcome:
\begin{itemize}
    \item The Gaussian eigenvalue distribution $p(\rho, \theta, \nu_e)$ is a poor description on small smoothing scales.
    \item At resolutions smaller than $\Senv$, the excursion set theory which underpins the first-crossing distribution is undefined.
    \item The $\nu\rightarrow\nu_{\rm eff}$ rescaling is known to be a poor approximation to the conditional mass function even in \text{$\Lambda$CDM}.
\end{itemize}
The first issue can be resolved in three ways. Following \cite{Alonso:2014zfa}, we could replace the functional form with the distribution of values from the {\it N}-body simulations.  We find this solution unsatisfactory, due to the requirement to run many realisations to avoid cosmic variance and the dependence upon individual simulation parameters.  However, as \cite{Alonso:2014zfa} note, this remains a useful technique to separate the effects of the Gaussian density field from the other approximations. Alternatively, it may be possible to add small-scale corrections to the probability density function, or to replace the Gaussian density field of \cite{1986ApJ...304...15B} with the log-normal density field proposed by \cite{Coles01011991}.

The second item appears to be insurmountable. This is because $M(\Senv)$ is completely controlled by our smoothing scale $\Renv$ via the window function.  Fortunately, on mildly non-linear scales and larger, there are so few haloes with $M > M(\Senv)$ that the problem is only an issue in mass bins which are already uncertain due to small-number statistics and cosmic variance.  Resolving these problems will improve the quantitative accuracy of our model beyond the large-scale, low-density regime. \cite{Alonso:2014zfa} found a better fit to the data by using an ``effective-universe'' approach, where halo abundances in an environment with overdensity $\denv$ are predicted as the unconditional abundance in a universe with effective cosmological parameters governed by the value of $\denv$, and the $\Senv$ mass cut is accounted for in an ad-hoc manner by limiting the Fourier modes available in that effective universe to those smaller than $\sim R_{\rm env}$. While there remain quantitative issues to overcome, our extremely simple model performs well qualitatively on all scales and in all density environments.

The last item is not readily solved, and is a common limitation of the conditional mass function. In order to produce an environment definition more akin to the excursion-set formalism we have also studied cases where the density field is filtered using a sharp-$k$ window function. As shown in \cref{fig:dfdlnm_renv_10_sharpk,fig:dfdlnm_renv_20_sharpk}, this produces a better agreement with the {\it N}-body simulations. A possible improvement may be to use an alternative to excursion set, \textit{e.g.} the survival probability approach of \cite{1990MNRAS.243..133P} or the Markovian Velocity excursion set of \cite{2014MNRAS.443.1601M}.

We have seen in the preceding two sections that the model illustrated in this paper is qualitatively useful but not always quantitatively accurate.  At large smoothing scales, where we are in the linear regime, our results agree well with the full {\it N}-body calculations.  As we move to smaller scales, we encounter the increasing non-Gaussianity of the environemnt overdensity and the chameleon regime of the $f(\R)$ modification.  While we can extend our model into this regime in low-density environments, it performs badly in knots, and by non-linear scales we only reproduce qualitative behaviour.  Nevertheless, we can attribute the difficulties to certain approximations and assumptions in our model, which in turn suggests avenues for improving our results.


\subsection{Multiplicity functions at fixed environment density} 
\label{sub:multiplicity_functions_at_fixed_environment_density}

\begin{figure}[!ht]
   \begin{center}%
        \subfloat[{$\Lambda$CDM, $\denv \in \left[ -0.5,-0.3 \right]$}]{\label{subfig:plot_dcut1_lcdm}\includegraphics[keepaspectratio,width=0.45\textwidth]{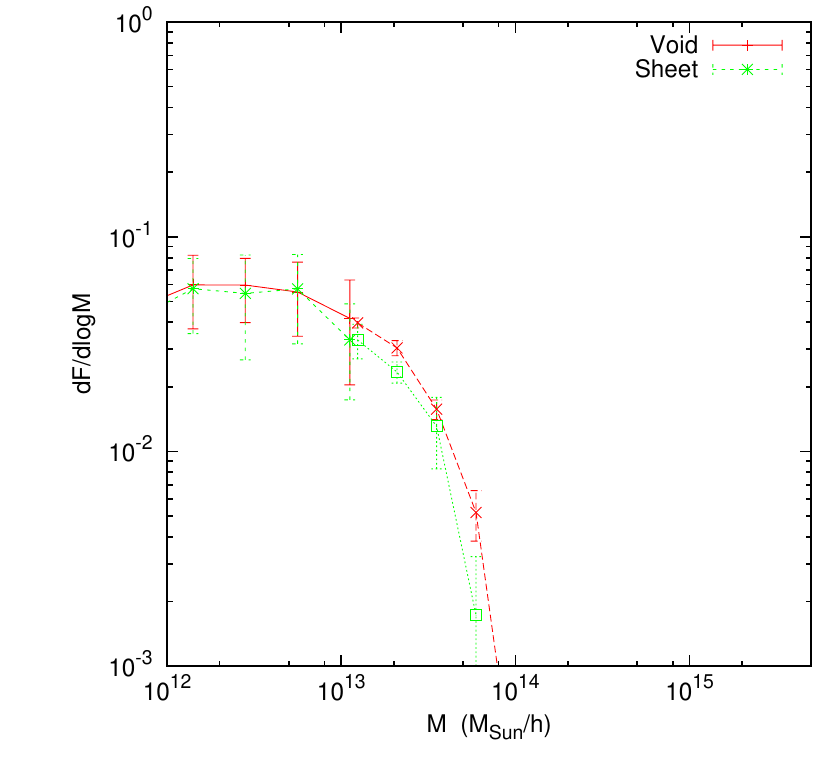}}\hfill
        \subfloat[{$\fr[5]$, $\denv \in [-0.5,-0.3]$}]{\label{subfig:plot_dcut1_f5}\includegraphics[keepaspectratio,width=0.45\textwidth]{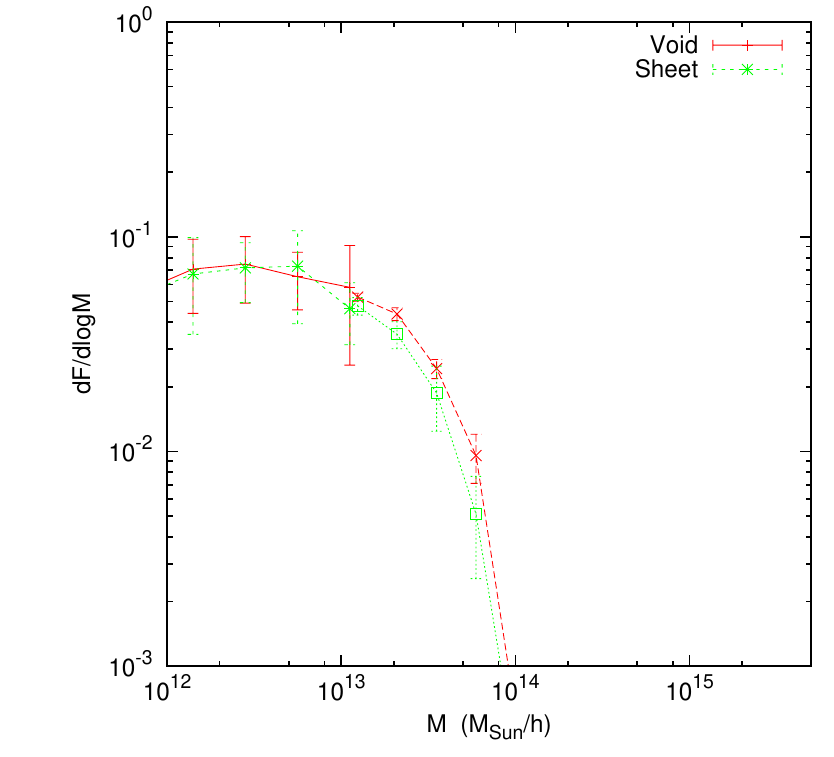}}\\
        \subfloat[{$\Lambda$CDM, $\denv \in [-0.1,0.1]$}]{\label{subfig:plot_dcut2_lcdm}\includegraphics[keepaspectratio,width=0.45\textwidth]{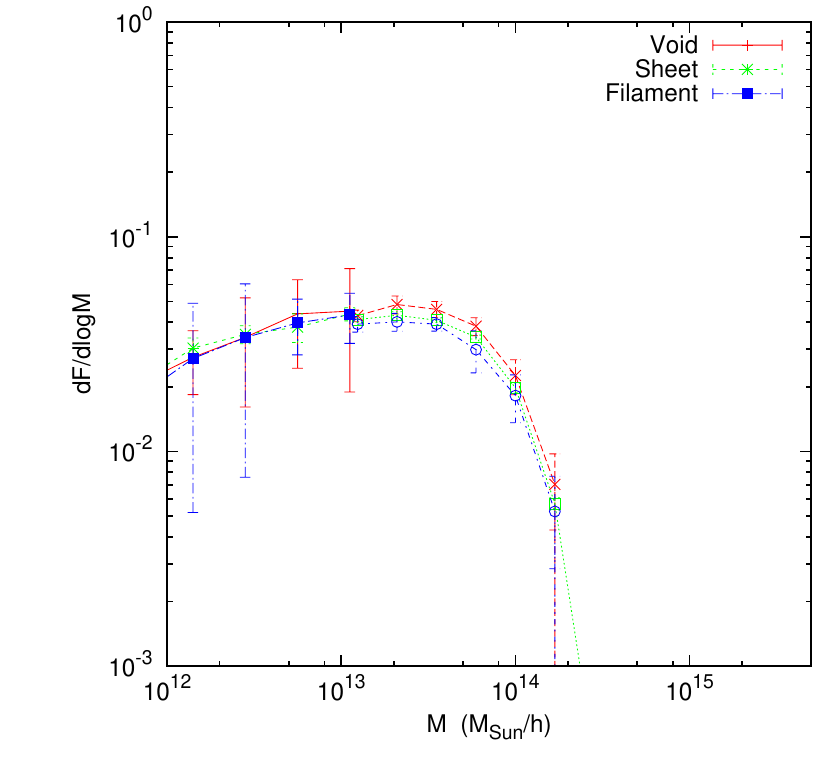}}\hfill
        \subfloat[{$\fr[5]$, $\denv \in [-0.1,0.1]$}]{\label{subfig:plot_dcut2_f5}\includegraphics[keepaspectratio,width=0.45\textwidth]{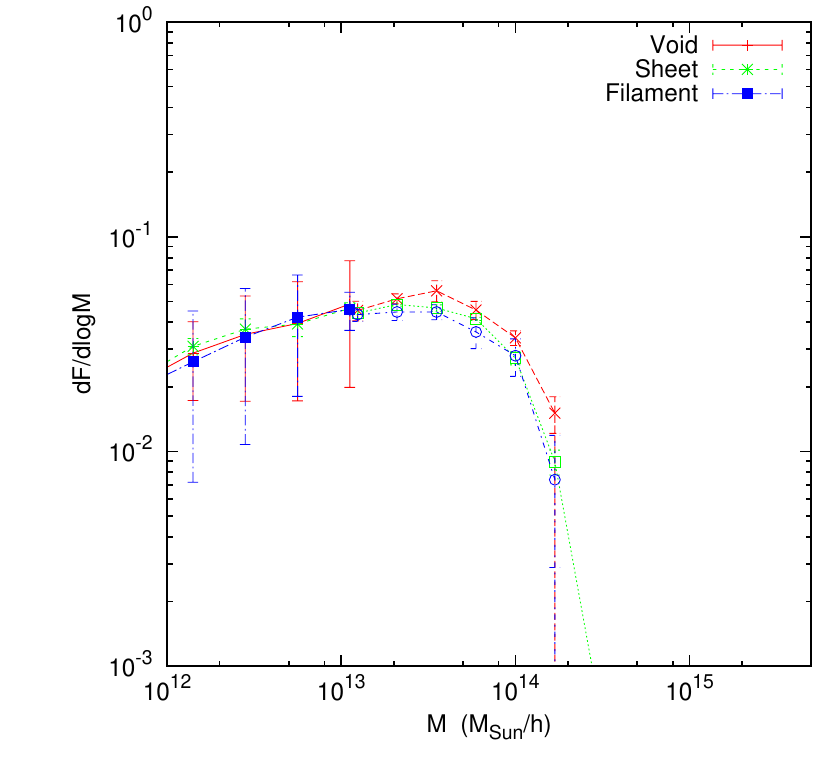}}\\
        \subfloat[{$\Lambda$CDM, $\denv \in [0.45,0.65]$}]{\label{subfig:plot_dcut3_lcdm}\includegraphics[keepaspectratio,width=0.45\textwidth]{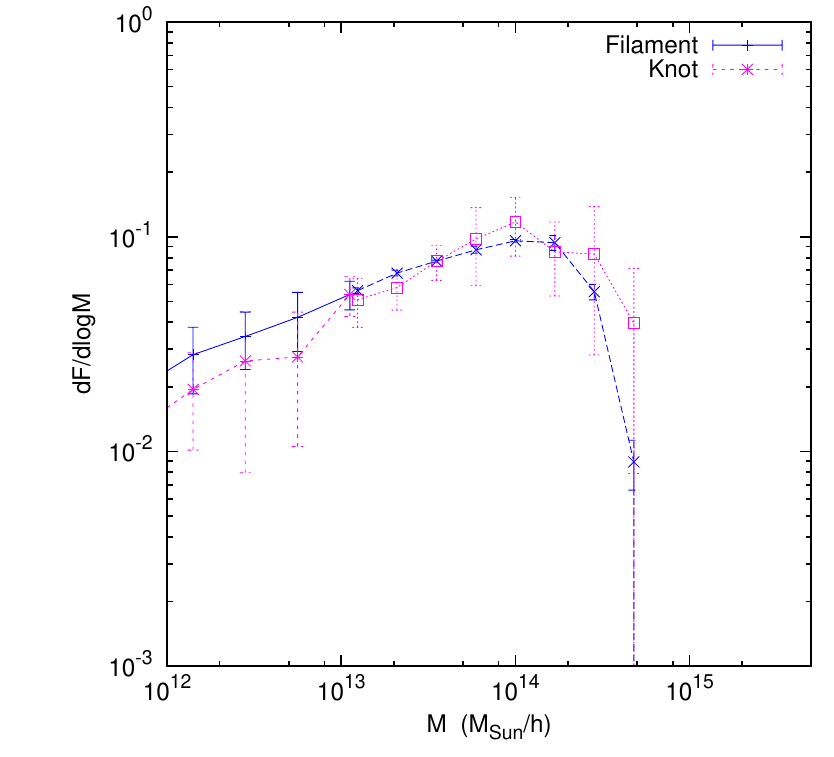}}\hfill
        \subfloat[{$\fr[5]$, $\denv \in [0.45,0.65]$}]{\label{subfig:plot_dcut3_f5}\includegraphics[keepaspectratio,width=0.45\textwidth]{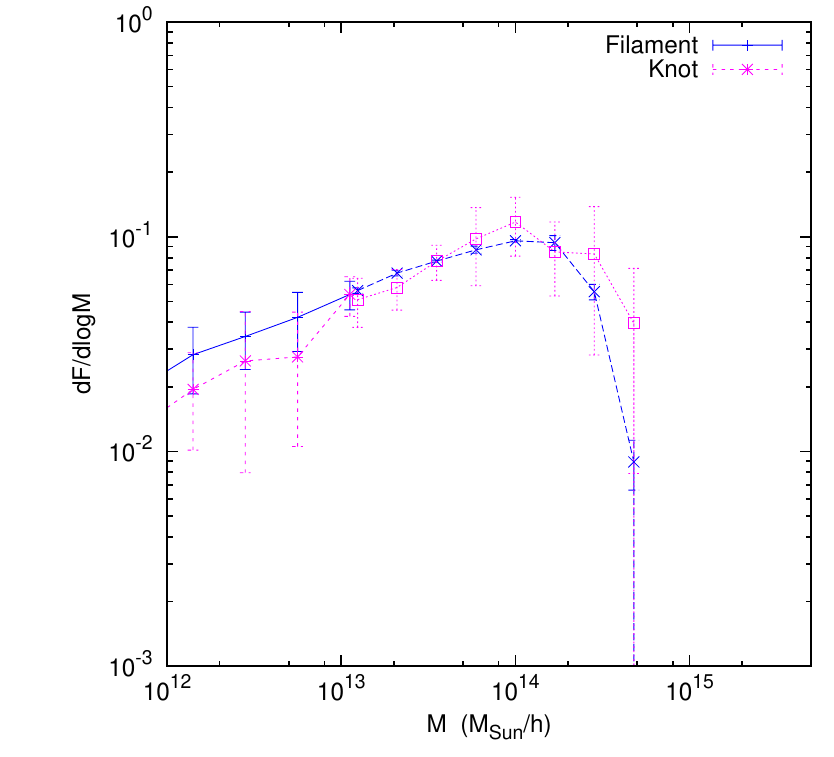}}%
   \end{center}\vspace{-1em}
    \caption{Multiplicity at $\Renv[10]$ after restricting the range of $\denv$.}\label{fig:densitymult}\vspace{-2em}
\end{figure}

One of the key predictions of the excursion-set formalism in $\Lambda$CDM cosmologies with Gaussian
initial conditions is that halo abundances only depend on the density of the environment, \textit{i.e.} not on
any other local environmental parameters. In spite of the approximate nature of this formalism, this
result was found to be true in \cite{Alonso:2014zfa} in comparison with {\it N}-body data. Therefore,
 it is interesting to explore whether the same is true for modified gravity
theories.

\cref{fig:densitymult} shows the conditional mass function for fixed environmental densities in different types of environments defined in terms of the eigenvalues of the tidal tensor.  We do not find significant differences between environments in terms of halo abundances. This is also a prediction of our semi-analytic model: we treat the environment as if it were $\Lambda$CDM (rather than with a full $f(\R)$ treatment) not only in calculating the environment density ODE, an approximation justified in \cite{Lombriser:2013wta}; but also (via the power spectrum) the environment variance.  Furthermore, the only change from $\Lambda$CDM to $f(\R)$ in our model is the barrier density, which itself is only a function of $\denv$ and not of $(\rho,\theta)$.  This is an \textit{a priori} simplification which is a corollary of only examining the modified gravity collapse assuming spherical symmetry, instead of the full ellipsoidal collapse. Nevertheless, as in the case of $\Lambda$CDM, we find that this phenomenological result is in agreement with the simulated data. It could be that $f(\R)$ does modify the $\Lambda$CDM result in certain regimes, but $\Renv[10]$ is a sufficiently large scale that the modifications are negligible, \textit{i.e.} the $f(\R)$ modification is in the cosmological regime where it must mimic the $\Lambda$CDM background.  Alternatively, perhaps the influence of $f(\R)$ is minimal on the structure of the cosmic web (\textit{e.g.} by late times non-linearities in the density field or the merger history of haloes may overwhelm any contribution from modified gravity) so the same physics dominates in both cosmologies. Regardless, the fact that we encounter the same result in both the semi-analytical and the numerical solution in encouraging.

\subsection{Internal halo properties} 
\label{sub:internal_halo_properties}

\begin{figure}
    \begin{center}
        \subfloat[$\Lambda$CDM]{\label{subfig:alignment_lcdm}\includegraphics[keepaspectratio,width=0.49\textwidth]{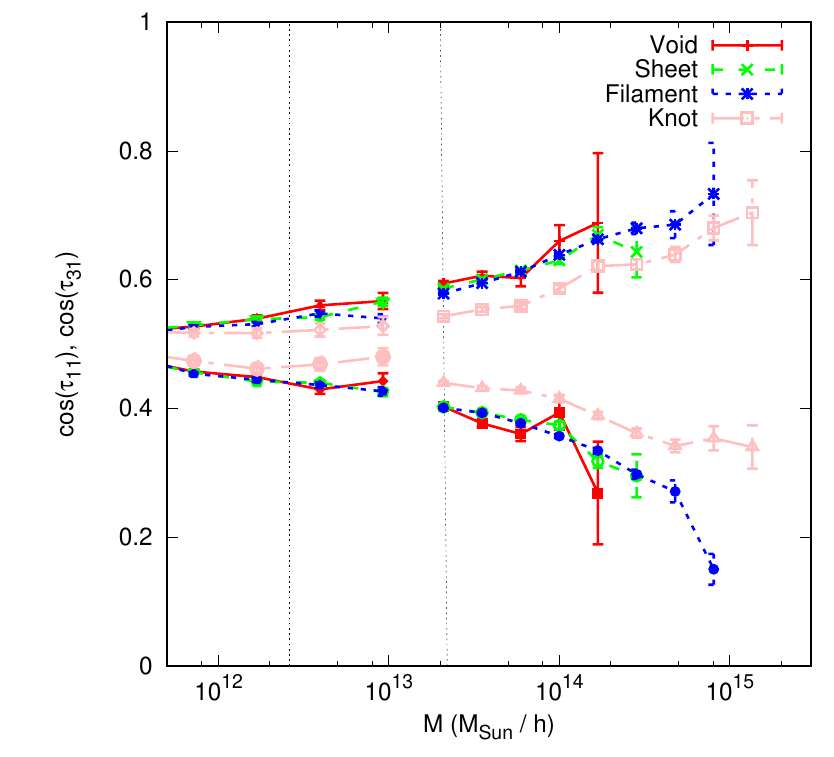}}
        \subfloat[{$\fr[5]$}]{\label{subfig:alignment_f5}\includegraphics[keepaspectratio,width=0.49\textwidth]{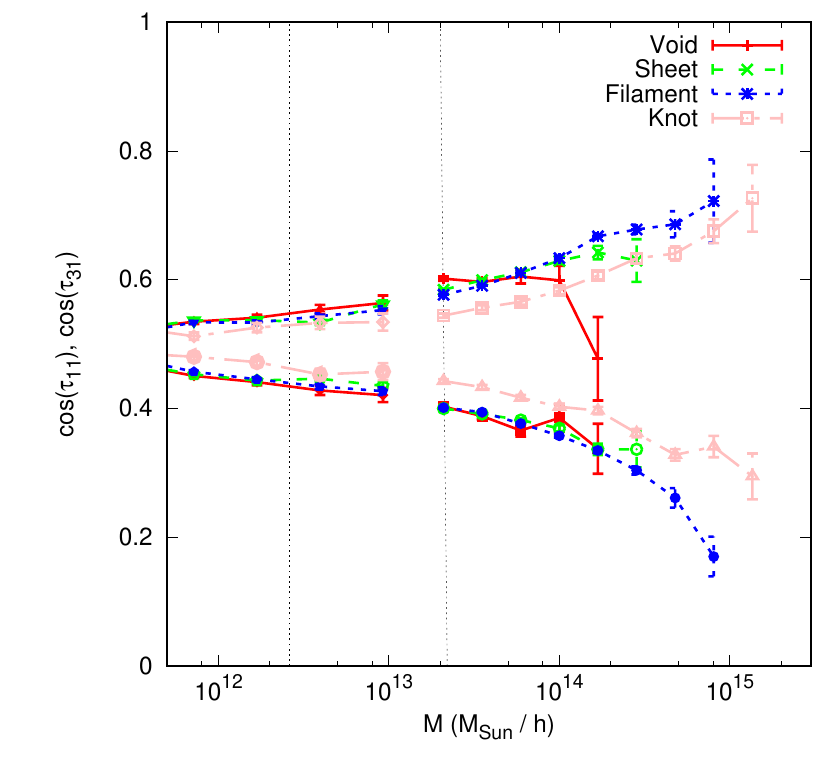}}
    \end{center}
    \caption{Alignment between the halos and the environment for $\Lambda$CDM (left) and $f(R)$ (right). The angle $\cos(\tau_{ij})$ is between the $i$-th axis of the halo and the $j$-th axis of the environment. The smoothing scale here is $R = 10~$Mpc$/h$ and eigenvalue threshold is $\lambda_{\rm th} = 0.1$. The vertical lines denote the mass for which we have $300$ particles in our halos for our two simulation boxes.}
\label{fig:alignment}
\end{figure}

\begin{figure}
    \begin{center}
        \subfloat[]{\label{subfig:spin_pdf_lcdm_f5}\includegraphics[keepaspectratio,width=0.49\textwidth]{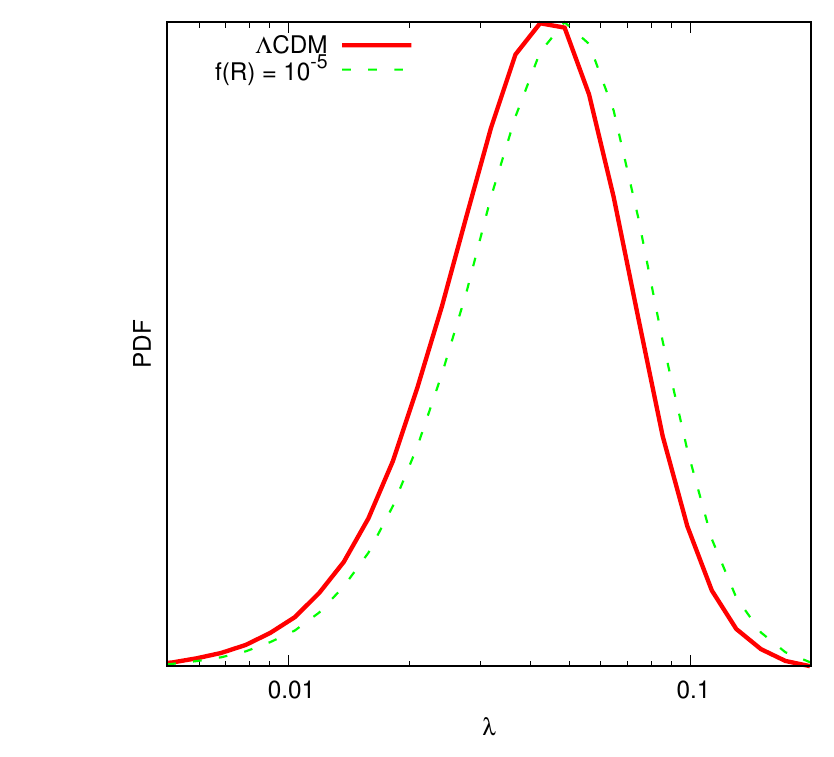}}
        \subfloat[]{\label{subfig:spin_pdf_ratio}\includegraphics[keepaspectratio,width=0.49\textwidth]{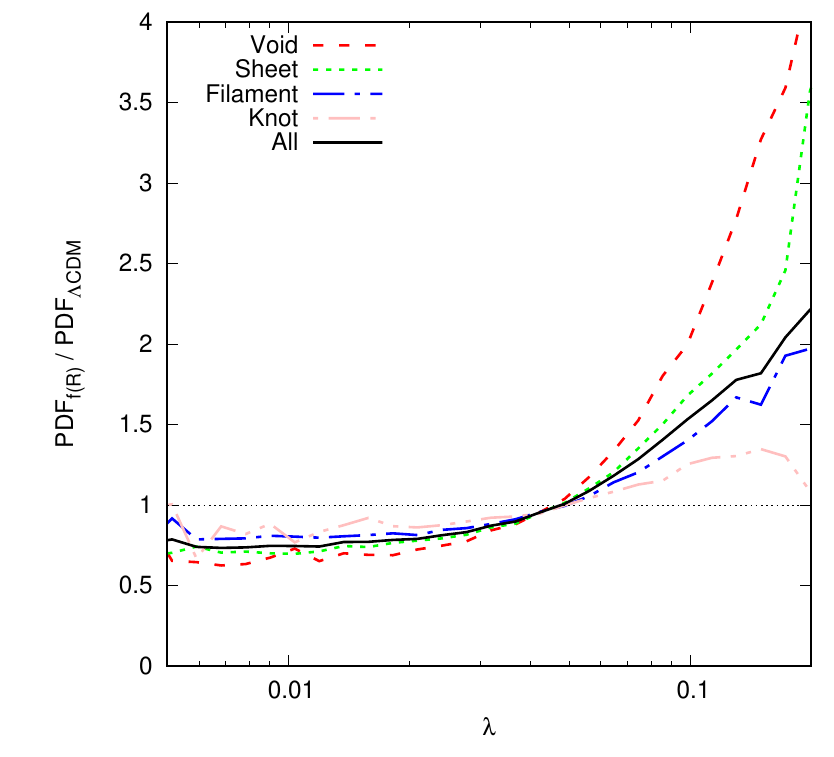}}\\
        \subfloat[]{\label{subfig:spin_pdf_lcdm_f5}\includegraphics[keepaspectratio,width=0.49\textwidth]{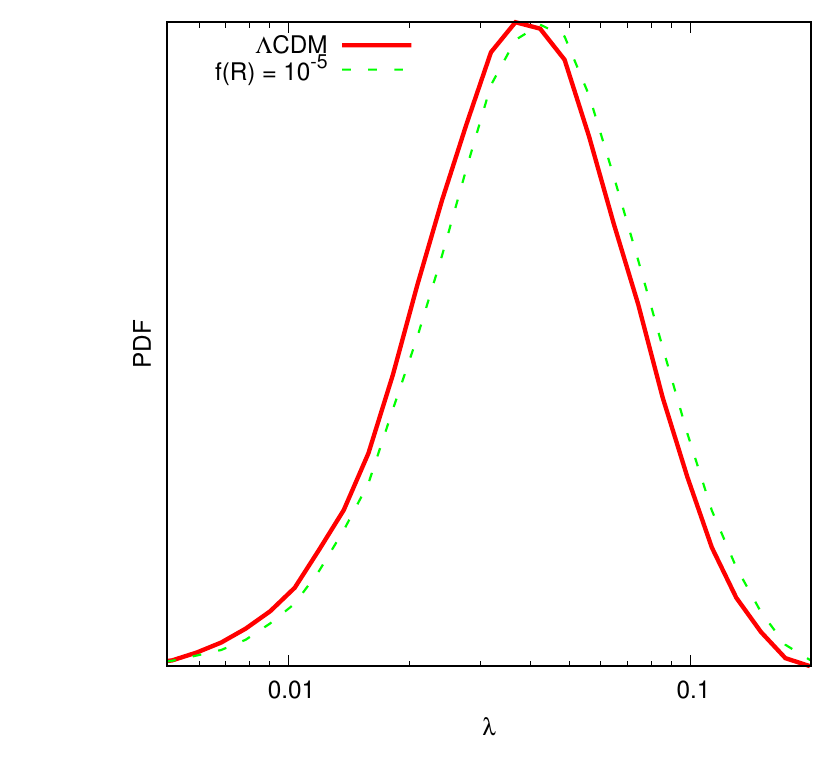}}
        \subfloat[]{\label{subfig:spin_pdf_ratio}\includegraphics[keepaspectratio,width=0.49\textwidth]{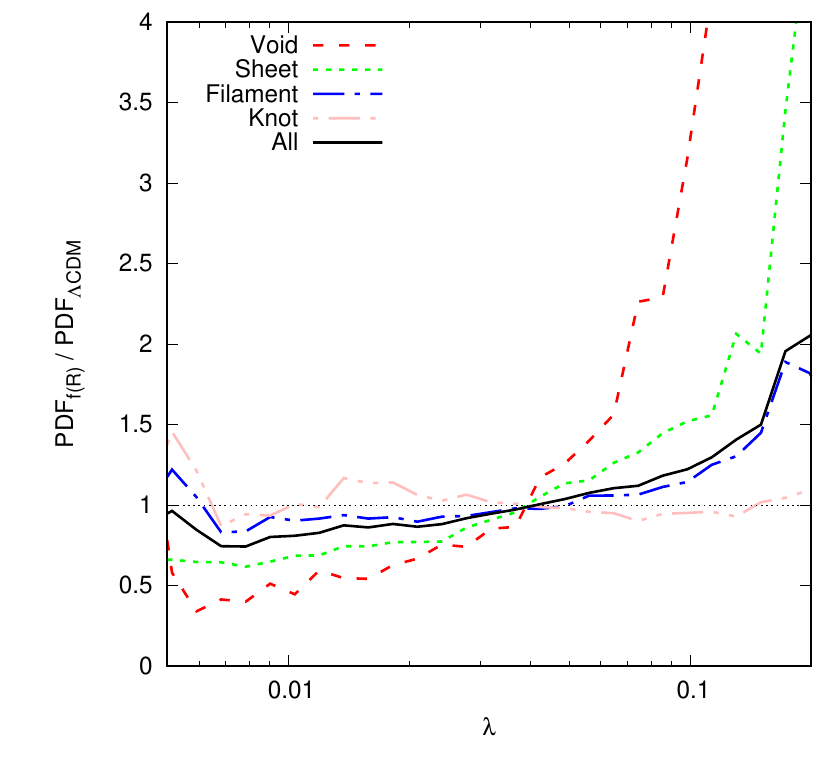}}
    \end{center}
    \caption{The PDF of spin-parameter (left) for $\Lambda$CDM and $f(R)$ in the $B = 1024$ Mpc$/h$ simulation. We also show the ratio of the PDF in $\Lambda$CDM to that in $f(R)$ for the four environments. The smoothing scale here is $R = 10~$Mpc$/h$ and eigenvalue threshold is $\lambda_{\rm th} = 0.1$. The upper (lower) panel shows the result of including halos which has more than $100$ ($300$) particles corresponding to only using halos with mass $>7\cdot 10^{12} M_{\odot}/h$ ($>2\cdot 10^{13} M_{\odot}/h$).}
\label{fig:spin}
\end{figure}

\begin{figure}
    \begin{center}
        \subfloat[]{\label{subfig:cnfw_lcdm_f5}\includegraphics[keepaspectratio,width=0.49\textwidth]{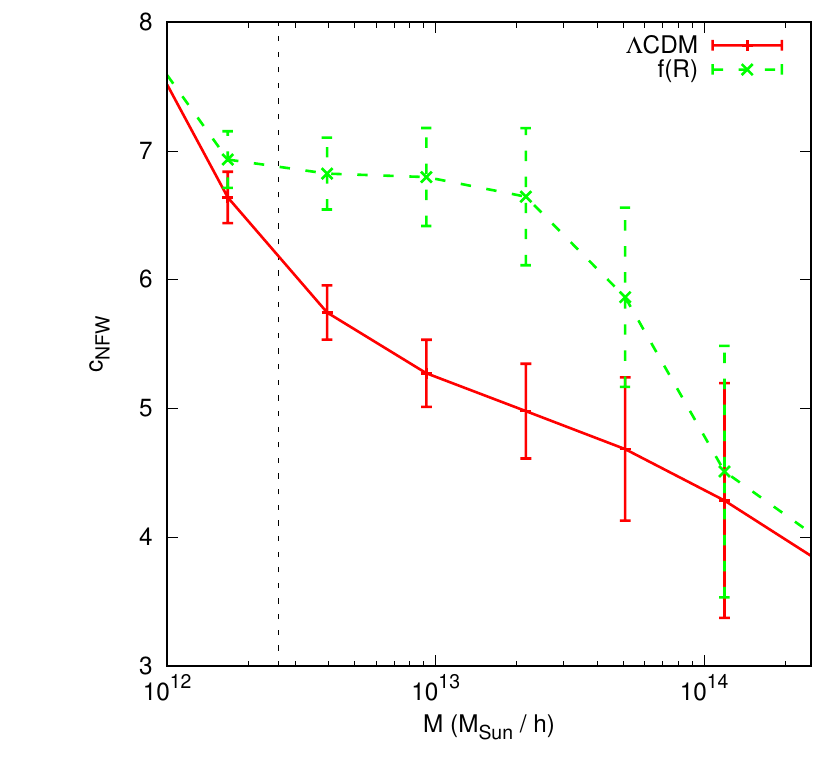}}
        \subfloat[]{\label{subfig:cnfw_ratio}\includegraphics[keepaspectratio,width=0.49\textwidth]{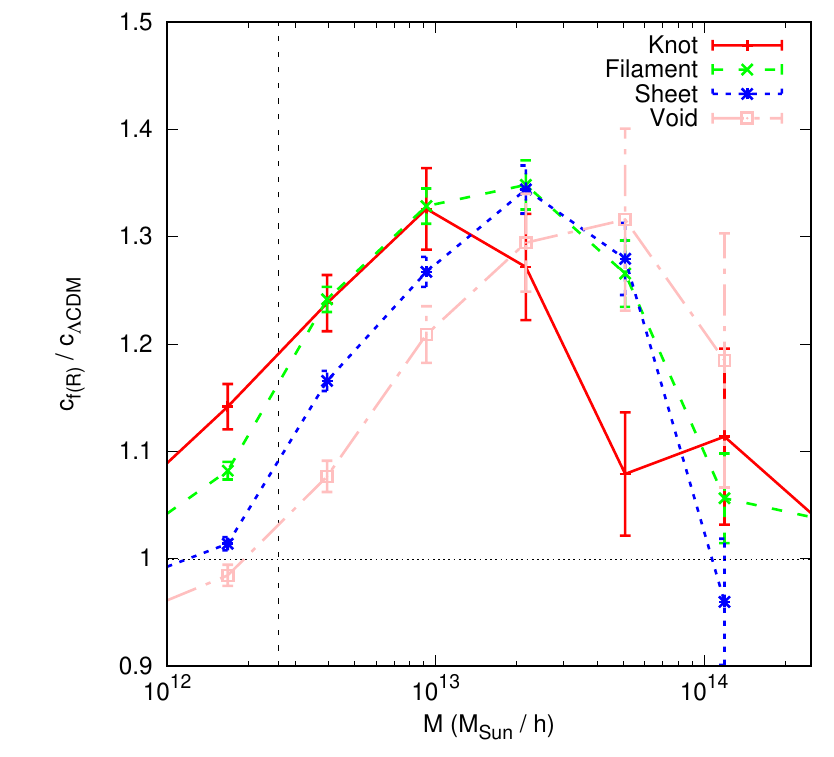}}
    \end{center}
    \caption{The NFW concentration parameter in $f(\R)$ and $\Lambda$CDM (left). We also show the ratio of the concentration parameter in $\Lambda$CDM to that in $f(\R)$ for the four environments (right). The smoothing scale here is $R = 10~$Mpc$/h$ and eigenvalue threshold is $\lambda_{\rm th} = 0.1$. The vertical line denotes the mass for which we have more than $300$ particles in our halos.}
\label{fig:cnfw}
\end{figure}

The relation of certain internal halo properties with the tidal structure of the environment could
also potentially contain relevant signatures for modified gravity models. To this end, we have studied the
relative orientation of the halo shapes and halo spins with respect to the eigenvectors of the
local tidal tensor, as well as the distribution of the halo spin parameter and the concentration-mass relation as a function of environment type. In order to obtain reliable estimates of these internal properties, we selected a subset of the halo catalog containing only virialized structures (as done in \cite{2014A&A...562A..78L}) with more than 100 particles per halo.  While \cite{2015MNRAS.454.2736C} suggest a cut-off of 300 particles per halo, on the basis of ensuring that the resolution bias is less than 10\% compared to the ellipticity result from sampling a halo of 1000 particles, our smaller cut-off produces a maximum resolution bias of $\sim 60\%$, which declines to $\sim 30\%$ as the ellipticity increases.  This affects the uncertainty on the shape of the haloes: thus, any difference between \lcdm{} and $\fr$ would have to exceed this uncertainty to be a non-null result.  We select only virialised structures by combining the kinetic energy $T$, gravitational potential energy $W$ and surface pressure term $E_{S}$ of the halo to define a parameter $\eta$ such that:
\begin{equation} \label{eq:virial_theorem}
	2T + W - E_S = 0 \implies \eta = \frac{2T - E_{S}}{W} + 1 \approx 0
\end{equation}
The first equality is the virial theorem, which defines a virialised structure by virtue of having reached a state of equilibrium \cite{2014A&A...562A...9G}.  The possibly non-zero value of $\eta$ allows for uncertainty in the measurement of the kinetic energy of the halo, so we set haloes with $\eta < 0.2$ to be virialised in accordance with the literature \cite{2014A&A...562A..78L}.  The decision to remove non-virialised haloes, rather than keeping all gravitationally-bound structures, stems from the different distributions of halo properties within the two classes of haloes \cite{2014A&A...562A...9G}.
We compute the relative orientation of the halo shape by estimating the inertia tensor of the halo as  \cite{ChisariZaldarriaga,2014A&A...562A..78L}:
\begin{equation}
  I_{ij} \propto \sum_{n=1}^{N_{\rm part}}x_{n,i}x_{n,j}
\end{equation}
where the position of the $n$-th halo is $x_{n}$ along each axis $i,j\in[1,3]$ and the sum is over all particles in the halo. The principal axis of the
halo is then defined as the eigenvector corresponding to the largest eigenvalue of $\hat{I}$, the (reduced) inertia tensor.
For each halo we then compute $\mu_{11}$ and $\mu_{13}$, the cosine of the angle between its
principal axis and ${\bf e}_1$, and ${\bf e}_3$, the eigenvectors of the local tidal field
(defined with a smoothing scale $R_{\rm env} = 10\,{\rm Mpc}/h$) corresponding to the largest
and smallest eigenvalues respectively. \cref{fig:alignment} shows the average $\mu_{11}$
and $\mu_{13}$ as a function of halo mass in the four different environment types. We find no
significant difference between $\Lambda$CDM and $f(\R)$ in the relative halo orientations
with respect to the cosmic web. We also performed a similar analysis studying the relative orientation between the principal directions of the cosmic web and the halo spin. We found no significant difference between $f(\R)$ and $\Lambda$CDM in this case either. Given the interpretation of \cite{2012MNRAS.427.3320C} that the alignment between dark matter haloes and the cosmic web is driven by the dynamics of cosmic flows during the merger history of the haloes, it is unlikely that this behaviour should be altered by modifications to gravity.  Specifically, the presence of the scalar field will affect the gravitational interaction during halo mergers, which are the key contributors to the final orientation and spin of the haloes.  However, if haloes flow preferentially towards increasingly collapsed structures, (\cite{2012MNRAS.427.3320C}; \textit{e.g.} along filaments towards knots), then the merger history of these haloes will occur in increasingly screened areas.  If the halo formed at early times (even in an unscreened void), each additional contribution to its orientation and/or spin vector will tend towards the $\Lambda$CDM equivalent, washing out any initial $f(\R)$ modification to the large-mass haloes.  In $\Lambda$CDM, \cite{2012MNRAS.427.3320C} find that haloes above a threshold of $M > (8 \pm 2) \cdot 10^{12} M_{\odot{}}$ exhibit this large-mass behaviour, which is the range we have tested here. Thus our null result is consistent with current interpretations of dark matter halo formation.

It is also customary to define the halo spin parameter \citep{2001MNRAS.321..559B}
\begin{equation} \label{eq:spin_lambda}
  \lambda\equiv\frac{|J|}{\sqrt{2G}M^{3/2}R^{1/2}},
\end{equation}
where ${\bf J}$ and $R$ are the halo angular momentum and virial radius respectively. The distribution of $\lambda$ is known to be only mildly dependent on mass\footnote{This dependence is described by \cite{2007MNRAS.376..215B} as \lq\lq{}small but real\rq\rq{} but the precise nature of it is so dependent upon the halo finder used that they do not quanitify the dependence for the FOF finder used in this paper.} and redshift in $\Lambda$CDM \cite{2007MNRAS.376..215B,2006MNRAS.370.1905H}, and therefore it is interesting to explore departures from this independence in modified gravity theories. \cref{subfig:spin_pdf_lcdm_f5} shows the spin distribution for all halos in the $\Lambda$CDM and $f(\R)$ simulations and \cref{subfig:spin_pdf_ratio} the ratio of the distributions between both cosmologies for the four different environment types.   The spin-parameter is seen to be higher in modified gravity which is in agreement with what was previously found in \cite{2013ApJ...763...28L}. We also split the halo catalog into two samples $M < 3\cdot 10^{13} M_{\odot}$ and $M > 3\cdot 10^{13} M_{\odot}$ and computed the PDF in each of the samples. For the high-mass sample the PDF is very close to $\Lambda$CDM while for the low-mass sample the PDF is enhanced in $f(\R)$ as shown in Figure \ref{fig:spin}. In this figure we also show how the result changes if we only include halos with more than $300$ particles ($M > 2.2 \cdot M_{\odot}/h$).

This agrees with \cite{2013ApJ...763...28L} where the conclusion (for the same $f(\R)$ model, but with $\abs{f_{\R 0}} = 10^{-6}$) was that modified gravity spins up galactic-sized halos.  We can intuitively understand why spin has to be modified via \cref{eq:spin_lambda}.  Let us approximate the spin as $\abs{\mathbf{J}} = M R \abs{\mathbf{v}}$.  Recalling \cref{eq:feff_newtonian}, we see that $G$ is enhanced by the factor $(1 + F_{\mathrm{eff}})$ with $F_{\mathrm{eff}} \in \left[ 0, \frac{1}{3} \right] $.  Thus we indirectly increase $\abs{\mathbf{v}}$, with $v \propto G$ in the linear regime, leading to the approximation $\lambda \propto \sqrt{(1 + F_{\mathrm{eff}})G}$.  In the absence of screening we would expect $\lambda^{f(\R)} / \lambda^{\Lambda \mathrm{CDM}} \approx \sqrt{1+\frac{4}{9}}$, a factor of $1.2$.  In practice, this is an upper bound because we have neglected the non-linear regime (in which $v\propto G$ does not hold) and the distribution of partial screening. The shift in the PDF we find is roughly $\lambda^{f(\R)} / \lambda^{\Lambda \mathrm{CDM}} \approx 1.04 $ which is within our upper bound. We also see a clear dependence of the spin-parameter in the cosmic web.  \cref{subfig:spin_pdf_ratio} demonstrates that the spin parameter is boosted mostly in low density environments (in voids and sheets) whereas in high-density environments (in filaments and knots) the value is close to $\Lambda$CDM for all values of $\lambda$.  This is as expected due to the environmental dependence of screening: a halo that is not screened if placed by itself might still be screened if placed in a high density environment.  Given the density ranges in \cref{eq:nueff_integral_limits}, the trend of enhancement with tidal morphology is what we expect from theory.

In Figure \ref{fig:cnfw} we see the dependence on the NFW concentration-mass parameter with mass and environment. The concentration-mass is enhanced in our $f(\R)$ model for halos with mass in the range $M \lesssim 10^{14} M_\odot / h$, whereas for halos of larger mass the results are close to $\Lambda$CDM.  This agrees with the results of \cite{2012PhRvD..85l4054L}, which only considered cluster-sized halos. There is some dependence on the cosmic web, but it is insignificant in comparison with the scatter in each mass-bin.

\section{Discussion} 
\label{sec:discussion}

We study the properties of dark matter haloes in the context of the environmental tidal classification of the cosmic web in $f(\R)$ theories. This classification defines four different types of environments based on the directionality of the tidal forces: {\sl voids} are defined as regions of space where tidal forces will expand an extended object in all directions, {\sl sheets} and {\sl filaments} have instead one or two compressing directions respectively, and in knots tidal forces will only compress structures. The relation between the gravitational potential and the density field also implies that these environment types will sample partially overlapping ranges of densities, and thus this classification can be a useful tool to study the effect of the chamaleon screening, present in viable $f(\R{})$ theories, on the properties of dark matter haloes.

We have described an approximate method to predict the abundance of dark matter haloes in each environment type.  Recall that our purpose was to use excursion set theory to take into account as much of the structure of the cosmic web and the non-linear collapse in $f(R)$ as possible in a semi-analytic model, in order to obtain results that can fit numerical simulations. Thus, we incorporated both the moving barrier of the excursion-set formalism---inherent in screened models---and the analytical description of the mass function conditional upon tidal environment described in \cite{Alonso:2014zfa}.  When comparing this prediction with the data from simulations we find that, even though the method is able to reproduce the correct behaviour of modified gravity theories qualitatively, the predictions become unreliable in high-density environment and for mildly non-linear filter scales. The method is, however, able to predict the enhanced abundance of high-mass objects found in low-density environments, a direct consequence of chamaleon screening, and the prediction is accurate in voids, where the departure from $\Lambda$CDM is largest. The shortcomings of this model are directly related to the inability of the traditional excursion-set formalism to accurately describe the conditional mass function, and we have outlined a number of ways forward to improve the predictions.

In spite of these shortcomings, we have gained further insight into the possible impact of modified gravity on halo properties conditional on environment by directly comparing the results of $f(\R)$ and $\Lambda$CDM simulations:
\begin{itemize}
 \item One of the key predictions in $\Lambda$CDM is that halo abundances should only depend on the enviromental density, and not on the directionality of the tidal forces. Our results show no significant departure from this behaviour in $f(\R)$, at least on mildly non-linear filter scales.
 \item We have identified a number of internal halo properties that are not affected by modified gravity in this context. In particular, we find that the alignment of halo orientations and spins with respect to the principal tidal directions of the environment.
 \item We have also seen that, even though the mass-concentration relation is different in $f(\R)$ and $\Lambda$CDM, its dependence on the tidal environment only shows marginally significant differences between both models.
 \item We observe a dependence in the distribution of the halo spin parameter upon tidal environment, with stronger deviations from the $\Lambda$CDM results in voids and sheets, as expected in the presence of chamaleon screening.
\end{itemize}

Incorporating morphological classification of the cosmic web into the existing analysis of dark matter halo properties increases their utility as a probe of modified gravity.  By doing so we can distinguish between the collapsed mass fraction in $f(\R)$ and $\Lambda$CDM far more than merely using the unconditional result, especially in low-density environments.  Furthermore, we can draw a boundary between the halo properties that are affected by screening and those which are not by explicitly examining the results in screened and unscreened regions, rather than averaging the effect over the distribution of environmental properties. While we find three null results, we also have three avenues for detecting the presence of $f(\R{})$ theories which are designed to evade local tests.

\section*{Acknowledgements}
We would like to thank Pedro Ferreira for many useful comments and discussions. HAW and DA are supported by the Beecroft Trust and the Oxford Martin School.

\bibliographystyle{JHEP}
\bibliography{bibliography}
\end{document}